\pdfoutput=1

\documentclass[letterpaper]{article}
\usepackage[T1]{fontenc}
\usepackage{url}

\usepackage{geometry}
\usepackage{setspace}
\usepackage[normalem]{ulem}
\usepackage{graphicx,color}
\usepackage{xcolor}
\usepackage{float}
\newfloat{scheme}{htbp}{los}
\floatname{scheme}{Scheme}
\floatname{chart}{Chart}
\newfloat{graph}{htbp}{loh}
\usepackage[articletitle = true, doi=true]{achemso}
\usepackage{booktabs}
\usepackage{longtable}
\usepackage{multirow}
\graphicspath{{figures/}}
\definecolor{teal}{HTML}{008080}
\usepackage[
    colorlinks=true,
	citecolor=teal,
	linkcolor=teal,
	urlcolor=teal]{hyperref}
\usepackage{url}
\usepackage{chemformula} % Formulas using \ch{}
\usepackage[version = 4]{mhchem} % Formulas using \ce{}

\newcommand{\device}[1]{{\texttt{ibm$\_$#1}}}

\newcommand{\ewf}[1]{EWF-#1}
\newcommand{\norb}{M}

\newcommand{\scinot}[2]{#1 \cdot 10^{#2}}

\usepackage{authblk}
\author[1]{Susanta Das}
\affil[1]{Computational Life Sciences, The Cleveland Clinic Research, The Cleveland Clinic, Cleveland, Ohio 44195, USA}
\author[2]{Thiago J. Pinheiro Dos Santos}
\affil[2]{Chemical Sciences Division,
Oak Ridge National Laboratory, Oak Ridge, TN, 37830 USA}
\affil[3] {National Center for Computational Sciences,
Oak Ridge National Laboratory, Oak Ridge, TN, 37830 USA}
\author[5]{Subhamoy Bhowmik}
\author[1]{Milana Bazayeva}
\author[1]{Zhen Li}
\author[1]{Akhil Shajan}
\author[1]{Danil Kaliakin}
\author[1]{Fangchun Liang}
\author[2]{Vyacheslav S. Bryantsev}
\author[3]{Al Geist}
\author[4]{Abigail McClain Gomez}
\author[4]{Thaddeus Pellegrini}
\author[4]{Robert Walkup}
\author[4]{Seetharami R. Seelam}
\author[4]{Mario Motta$^\S$}
\author[1,5]{Kenneth M. Merz, Jr.$^\dagger$}
\author[3]{Thomas Beck$^\ddagger$}
\affil[4]{IBM Quantum, IBM T.J. Watson Research Center, Yorktown Heights, NY 10598, USA}
\affil[5]{Department of Chemistry, Michigan State University, East Lansing, Michigan 48824, USA}

\date{$^\S$mario.motta@ibm.com, $^\dagger$merzk@ccf.org, $^\ddagger$becktl@ornl.gov}

\title{Quantum Computations on Fusion Blanket Molten Salts\footnote{Notice: This manuscript has been authored by UT-Battelle, LLC, under contract DE-AC05-00OR22725 with the US Department of Energy (DOE). The US government retains and the publisher, by accepting the article for publication, acknowledges that the US government retains a nonexclusive, paid-up, irrevocable, worldwide license to publish or reproduce the published form of this manuscript, or allow others to do so, for US government purposes. DOE will provide public access to these results of federally sponsored research in accordance with the DOE Public Access Plan (https://www.energy.gov/doe-public-access-plan).}}

\begin{document}

\maketitle

\begin{abstract}
Molten salts such as FLiBe (2LiF--BeF$_2$) are leading blanket materials for breeding and recovering tritium in fusion reactors. Predicting tritium speciation requires accurate electronic ground-state energies for representative molten-salt clusters, a demanding task for correlated electronic-structure methods. Here we report the first application of heterogeneous quantum--classical computing to tritium binding in FLiBe. Clusters drawn from \emph{ab initio} molecular dynamics are partitioned by an embedded-wavefunction (EWF) method into atom-centered fragments, and the largest fragments are solved on IBM quantum hardware using extended sample-based quantum diagonalization (ext-SQD). Across nine clusters, the heterogeneous quantum--classical workflow reproduces fragment ground-state energies with agreement to full configuration interaction within 0.7~kcal/mol and a mean absolute deviation of 0.3~kcal/mol. In contrast, fragmented and unfragmented conformational energy differences and tritium binding energies differ by 12~kcal/mol and 110~kcal/mol on average, respectively, identifying fragment construction rather than fragment solution as the dominant source of algorithmic bias. To the best of our knowledge, this is the first such demonstration for a charged ionic system and in particular an inorganic molten salt, where electrostatic and polarization effects make the accurate treatment of electronic correlation particularly challenging. These results also identify areas of future research towards an accurate and scalable quantum--classical workflow to compute free-energy estimates of tritium speciation in fusion blankets.
\end{abstract}

Nuclear fusion has great potential to provide base-load electrical energy that can meet the challenge of the world's growing energy demands. \cite{doe_fusion_roadmap} There has been a recent rapid growth in efforts to develop net-energy-producing fusion reactors, both in the commercial and government-funded sectors.\cite{anderson2025comprehensive,hillesheim2026overview}  

Two overarching science and engineering challenges face the fusion community:\cite{doe_fusion_roadmap} 1) high energy-density plasmas that are stable over times long enough to carry out the deuterium/tritium (D/T) fusion reactions that produce heat-generating radiation and 2) sufficient tritium fuel production (breeding) in a blanket material surrounding the plasma core.  Recovering the tritium produced in the blanket material for re-insertion into the plasma is an essential challenge to be addressed.  A 1 GW reactor will require roughly 0.5 kg of tritium per day, a significant fraction of the current global stockpile of tritium totaling $\approx 25$ kg. 

At the computational level, large-scale magnetohydrodynamic (MHD) simulations have guided the development of reactor geometries and magnetic field designs to confine the plasma. \cite{staebler2022advances} The blanket materials challenge involves both macroscopic MHD models of flows and atomic-level models that aim to reproduce and predict properties of proposed materials that optimize both tritium production and recovery.\cite{clark2025breeder} Here quantum mechanical simulations will play a crucial role.  

Three candidates have been proposed for blanket materials surrounding the plasma.\cite{clark2025breeder}  All three include the $^6$Li isotope, which, when hit by an energetic neutron produced by fusion in the plasma, fragments into a $^4$He atom and a tritium ion (each carrying significant kinetic energy). The proposed materials are: 1) Lithium containing ceramics (solid breeders), 2) liquid metals such as a lead/lithium mixture, and 3) molten salts, with FLiBe the primary candidate \cite{hillesheim2026overview}. 

In this work, we focus on tritium speciation in a FLiBe molten salt blanket. We make this choice  because FLiBe has several advantages for high-magnetic-field, high-energy-density fusion reactors.\cite{hillesheim2026overview} FLiBe is also one of the leading materials in small modular fission reactors.\cite{zhang2018redox,jiang2025solute}   

Here we provide the first benchmark results that apply quantum computing to tritium binding in FLiBe. The initial aim is to assess our ability to calculate accurate ground-state energies of tritium binding in clusters of the molten salt FLiBe.
These systems present significant challenges to quantum calculations due to the ionic character of FLiBe. Negatively charged fluorine F$^-$ has large polarizability, with dynamic correlations that produce long-range dispersion interactions. Positively charged beryllium Be$^{2+}$ interacts intensely with fluorine, forming tetrahedral structures, and leading to a fluctuating structural order in the liquid\cite{lam2021modeling}. Tritium tends to bind strongly to one or more fluorine atoms, often forming bent/stretched multi-center F-T-F bonds. Bonding in isolated F-T-F anions is itself a complex and interesting subject\cite{dunning2021bifluoride}, made even richer by the presence of the FLiBe environment.

Previous quantum chemical studies of tritium in molten salts have been performed at the density functional theory (DFT) level \cite{nam2015redox,lam2021impact,wang2022first} or with machine learning force fields (MLFF) derived from DFT simulations. \cite{lam2021modeling,porter2022computational,jiang2025solute} Lam et al.\cite{lam2021impact} performed \textit{ab initio} molecular dynamics (AIMD) DFT simulations of tritium in FLiBe and FLiNaK molten salts. In the FLiBe salt, tritium diffuses much faster in the T$_2$ (neutral) form compared to the T$^+$ (ionic) form.  This is due to weaker interactions of the molten salt with the neutral molecule.  The differences are clearly reflected in the radial distribution functions that illustrate the weaker binding of the neutral species.  Similar behavior is observed in the FLiNaK salt.  

Wang et al.\cite{wang2022first} used AIMD simulations to study the basic structural and thermodynamic properties of FLiBe, which were in good agreement with the experiment. The authors observed structural properties similar to those seen by Lam et al., with the tritium ion tightly bound to fluoride ions, while the neutral atom or the T$_2$ molecule interact weakly.  

In a recent study, Jiang et al.\cite{jiang2025solute} performed extensive ML simulations, built on AIMD forces and energies, to examine the impact of various solutes on the tritium redox chemistry. It was found that reducing agents such as the Be atom can convert ionic tritium to the neutral species. Experimental research is also being devoted to the possibility of chemical control. \cite{zhang2018redox} Further exploration of this strategy for impacting tritium speciation should be explored, and will be a goal of our future work. 

Finally, Shi, Lam, and Beck\cite{shi2022deep} performed the first DFT-AIMD/MLFF estimation of the thermodynamic chemical potentials of the various species in a molten salt (NaCl).  Quasi-chemical theory was used to compute the chemical potentials.  From the excess chemical potentials, it is easy to construct the excess Gibbs free energy for the liquid.  The authors found that the DFT-level models deviated from the experimental free energy by roughly 10\%, an error level that calls into question the predictive capability of DFT for computing tritium affinities.  
A correction scheme was then developed in which instantaneous clusters were extracted from the simulations, and quantum chemical (MP2-level) computations were performed to refine the free energy estimation.  This improved the accuracy to less than 1\% error in the free energy. 

The goal of the present study, as a first step, is to use state-of-the-art quantum devices to assess our ability to obtain accurate results for tritium binding energies that will subsequently feed into the free energy calculations. Because excess chemical potentials drive the chemical phenomena of interest, we require high-fidelity calculations to obtain predictive results. See the Supporting Information for discussion of the free energy calculations into which the quantum binding energy calculations will be fed in future studies.  

\begin{figure}[htbp]
  \centering
  \includegraphics[width=0.9\linewidth]{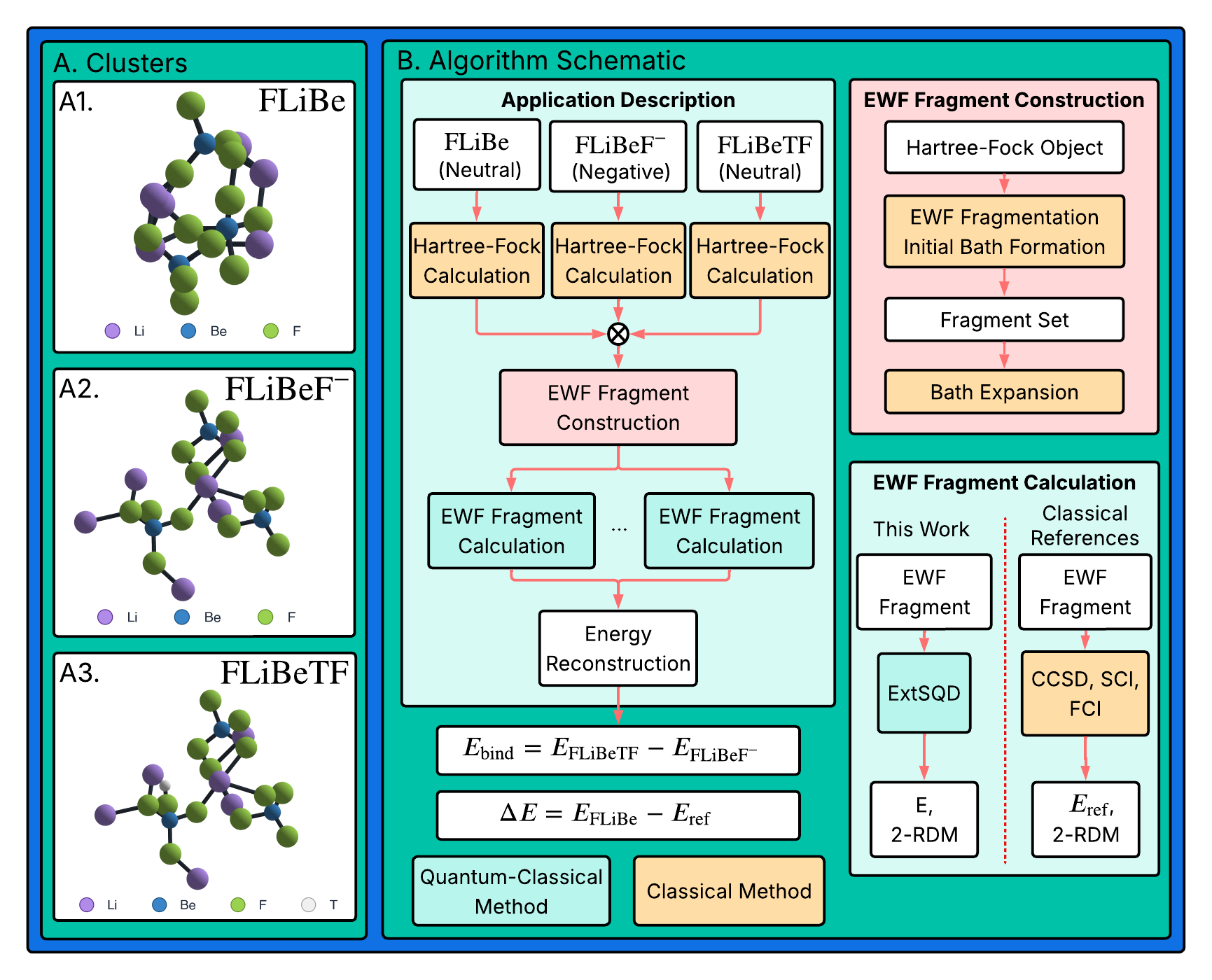}
  \caption{High-level overview of the heterogeneous quantum--classical embedded-wavefunction (EWF) workflow used in this work. (A)~Representative cluster conformations: (A1)~21-atom \ce{Li6Be3F12} (FLiBe), (A2)~22-atom \ce{[Li6Be3F13]-} (FLiBe$^-$), and (A3)~23-atom \ce{Li6Be3F13T} (FLiBe$+$T). Color code: Li (violet), Be (blue), F (green), T (grey). (B)~Each conformation of each cluster is described by a Hartree--Fock calculation that feeds the EWF fragment construction, in which the Hartree--Fock object is partitioned into atom-centered fragments and a fragment Hamiltonian is built for each through localized-orbital fragmentation and localized two-electron-integral construction. Every fragment is then solved either by the quantum--classical route discussed in Fig.~\ref{fig:sqdmethods}, extended sample-based quantum diagonalization (ext-SQD), or by classical reference solvers (CCSD, truncated CI, and FCI); each solver returns a fragment energy and two-body reduced density matrix, which are recombined in the energy-reconstruction step to give the total energy. For the tritium-binding application the target observable is $E_{\mathrm{bind}} = E_{\mathrm{FLiBeTF}} - E_{\mathrm{FLiBeF}^-}$}.
  \label{fig:method}
\end{figure}

The computational strategy used in this study is illustrated in Fig.~\ref{fig:method} which comprises problem fragmentation based on the Embedded Wave Function (EWF) method, and quantum-mechanical solution of fragments by the extended sample-based quantum diagonalization (ext-SQD) method, that we now briefly describe.

\textbf{Embedded Wave Function (EWF).} 
The conformations (i.e. atomic geometries) of the FLiBe clusters treated in this work are obtained from DFT-AIMD/MLFF trajectories of pristine or tritiated FLiBe melts, by extracting instantaneous configurations and subselecting groups of 21-23 atoms (as detailed in the Supporting Information). These clusters play the same role as the clusters in the quantum-chemical correction step of the quasi-chemical free-energy framework described above~\cite{shi2022deep}, and the accuracy of their computed energies sets the accuracy of that correction. We treat electron correlation in each conformation of each cluster with the embedded wave-function (EWF) method\cite{nusspickel2022systematic} (Figure~\ref{fig:method}), a quantum fragmentation scheme rooted in density-matrix embedding theory (DMET).\cite{knizia2012density} Within EWF, a mean-field reference fixes a set of localized fragments, each consisting of a single atom together with bath orbitals that encode the one-particle entanglement between that atom and its environment. Every fragment is solved once with a correlated solver, and its one- and two-particle density matrices are mapped back onto the full molecular orbital space. Summing these projected contributions over all fragments yields the global observables of interest, particularly the total energy, without requiring any self-consistency loops or matching condition between fragments.

For the present %\ce{Li6Be3F12} 
clusters, a restricted Hartree--Fock calculation using the 6-31+G(d) basis provides the reference one-particle density matrix, from which Intrinsic Atomic Orbitals (IAOs)\cite{knizia2013intrinsic} are constructed to define a localized and chemically meaningful basis for the occupied space. This choice is well suited to FLiBe, where bonding ranges from highly ionic Li--F to partly covalent Be--F, and where an atom-centered partition with atom-environment entanglement captured through a standard 
DMET Schmidt decomposition of the mean-field density matrix gives a reasonable chemical picture of the bonding in each conformation. The same atom-centered description carries over directly to the tritiated and anionic clusters of System~2: the tritium binds as a bare \ce{T+} nucleus held electrostatically by fluoride, and the excess-electron fluoride of the \ce{[Li6Be3F13]-} anion adds further polarization, so both species remain ionic, atom-centered problems for which the IAO partition and DMET bath are equally well suited.

The compact DMET entanglement bath obtained in this way is then expanded in a manner analogous to pair natural orbitals. Additional occupied and virtual states are introduced from an MP2 amplitude analysis around each atom, and natural orbitals with occupations larger than a threshold $\eta = 10^{-5}$ (for virtuals) or smaller than $2 - \eta$ (for occupieds) are retained, systematically recovering the local excitations needed for an accurate treatment of local electronic correlation. This MP2-bath expansion yields fragments whose orbital count ranges from 8--15 molecular orbitals for the Li- and Be-centered fragments to 27--33 molecular orbitals for the fluorine-centered fragments. The larger fluorine spaces reflect the polarization of the fluoride lone pairs and their mixing with the neighboring Be 2p orbitals, which demand a richer correlated description. In the tritiated System~2 clusters the tritium-centered fragment is comparably compact, comprising 12 molecular orbitals on average (one fragment orbital plus MP2-augmented bath; range 10--15 across the nine clusters).

All EWF calculations are performed with the Vayesta package\cite{nusspickel2022systematic,nusspickel2023effective}, and total energies are reconstructed from the projected fragment density matrices using the partitioned cumulative expression for the correlation energy on top of the global Hartree--Fock reference~\cite{nusspickel2023effective}. Within this protocol, we solve every fragment classically at both the coupled-cluster singles and doubles (CCSD) and the full configuration interaction (FCI) levels, giving \ewf{CCSD} and \ewf{FCI} total energies. The heterogeneous quantum--classical total energy \ewf{FCI+ext-SQD} is built from the same fragment density matrices: the classical FCI reduced density matrices are retained for the small fragments ($\norb < 13$), and the ext-SQD reduced density matrices obtained on the IBM quantum processor are substituted for the larger fragments ($\norb \ge 13$). Fragmentation, bath construction, and partitioned-cumulant reconstruction are kept fixed across all three EWF totals, so any difference between \ewf{CCSD}, \ewf{FCI}, and \ewf{FCI + ext-SQD} comes from the fragment solver alone. The gap between this family of EWF calculations and full-molecule methods measures the error from EWF fragmentation.

The same EWF workflow, with the same IAO localization, MP2-augmented DMET bath, and partitioned-cumulant assembly, has recently been deployed for protein-scale quantum-chemical embedding\cite{shajan2025molecular}. Its transfer to conformations of pristine and tritiated FLiBe clusters extracted from finite-temperature AIMD snapshots shows the versatility of the EWF-based approach.

\begin{figure}[htbp]
  \centering
  \includegraphics[width=\linewidth]{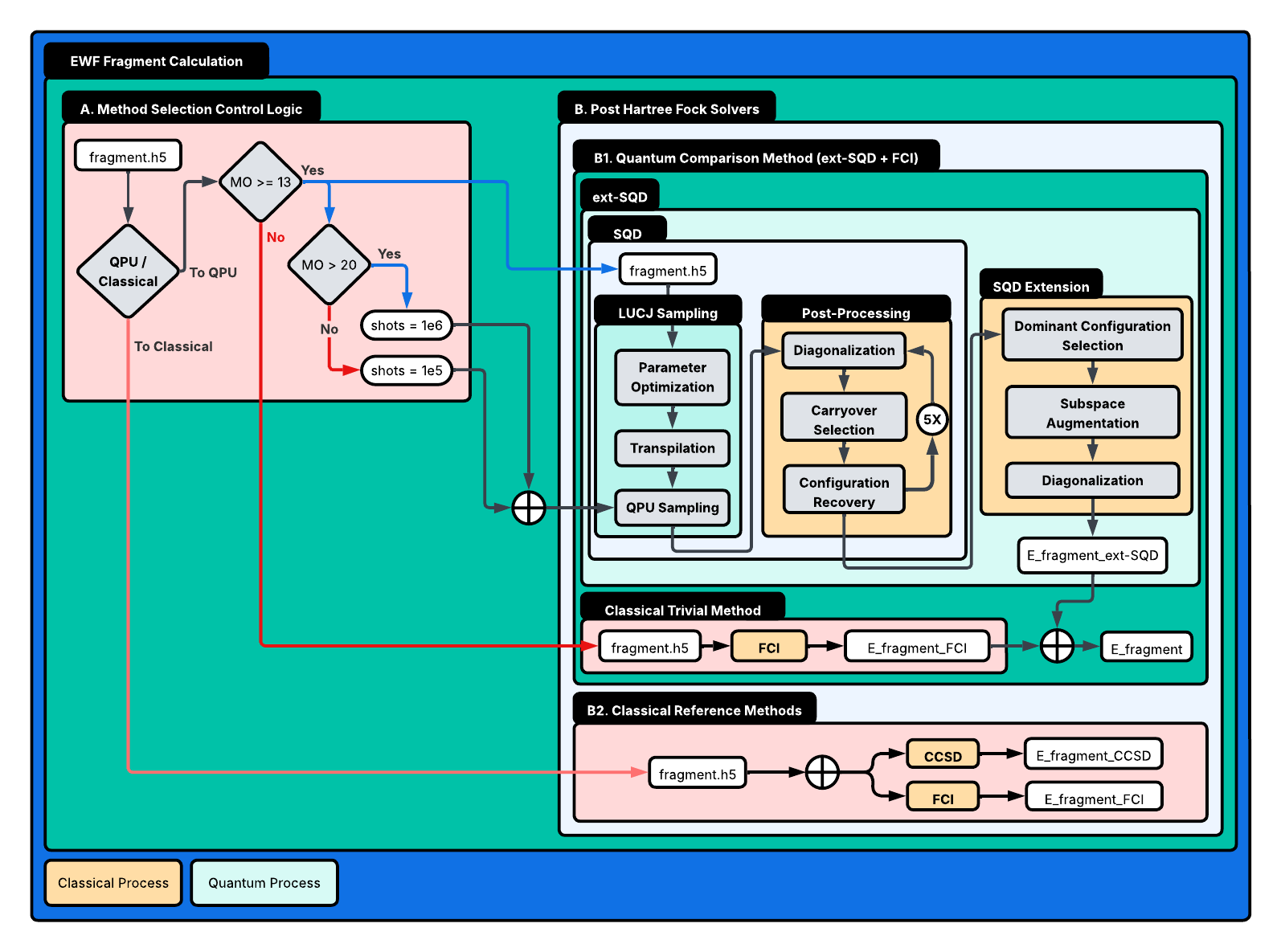}
  \caption{Schematics of EWF fragment calculations. Every fragment Hamiltonian with 13 or more spatial orbitals enters the ext-SQD pipeline: quantum sampling --- LUCJ ansatz preparation, transpilation, and circuit execution on the IBM Heron-r3 processor \device{boston} --- generates candidate electron configurations; the subspace-diagonalization loop (configuration recovery, exact diagonalization of the projected Hamiltonian, and carryover selection of the dominant configurations) is iterated five times; the converged subspace is then extended through single electronic excitations (subspace filtering, expansion, and a final diagonalization), yielding the ext-SQD fragment energy and two-body reduced density matrix that are returned to the EWF energy reconstruction. Green (orange) boxes denote quantum (classical) processes.}
  \label{fig:sqdmethods}
\end{figure}

\textbf{Sample-based Quantum Diagonalization (SQD) and ext-SQD.} For the larger fragments, classical FCI is feasible for a one-off check but too slow to run across a full conformational campaign. For these fragments, we replace the fragment solver by sample-based quantum diagonalization (SQD)~\cite{robledo2025chemistry,motta2024subspace,lassqd,kanno2026quantum} (Figure~\ref{fig:sqdmethods}). In SQD, candidate electron configurations are drawn from a parameterized quantum state prepared on a quantum processor, a configuration recovery operation is used to mitigate errors, and a compact CI Hamiltonian is then diagonalized classically in the resulting subspace. The quantum device thus acts as a configuration generator, while classical post-processing recovers the fragment ground-state energy and reduced density matrices.

\begin{figure*}[htbp]
\centering
\includegraphics[width=0.55\textwidth]{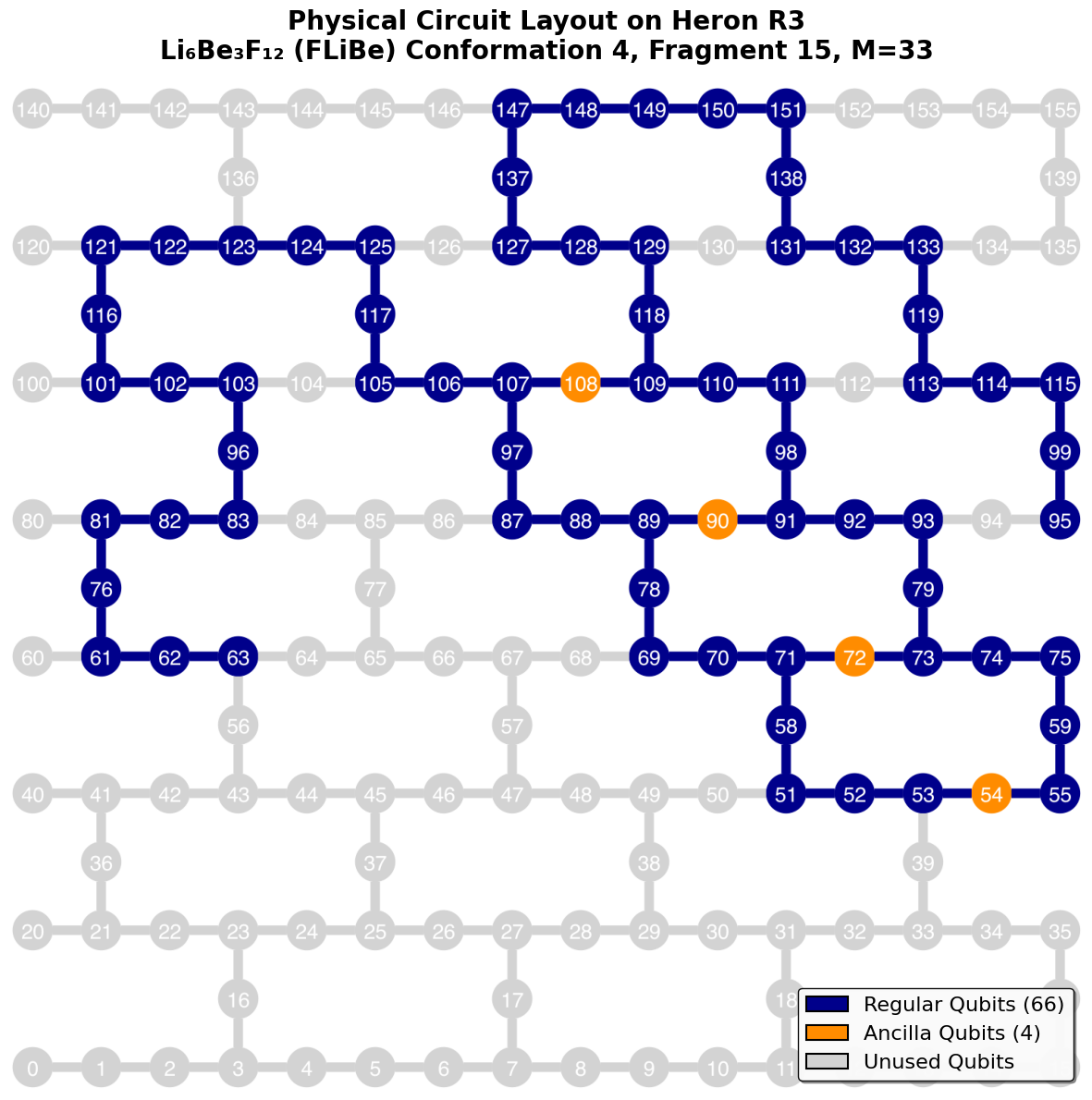}
\caption{Physical circuit layout on Heron R3 of the largest fragment, corresponding to cluster 4, fragment 15, with $\norb = 33$. Main-register qubits are shown in dark blue, ancilla qubits in orange, and unused qubits in grey.}
\label{fig:largest_fragment_layout}
\end{figure*}

\begin{figure*}[htbp]
\centering
\includegraphics[width=1.0\textwidth]{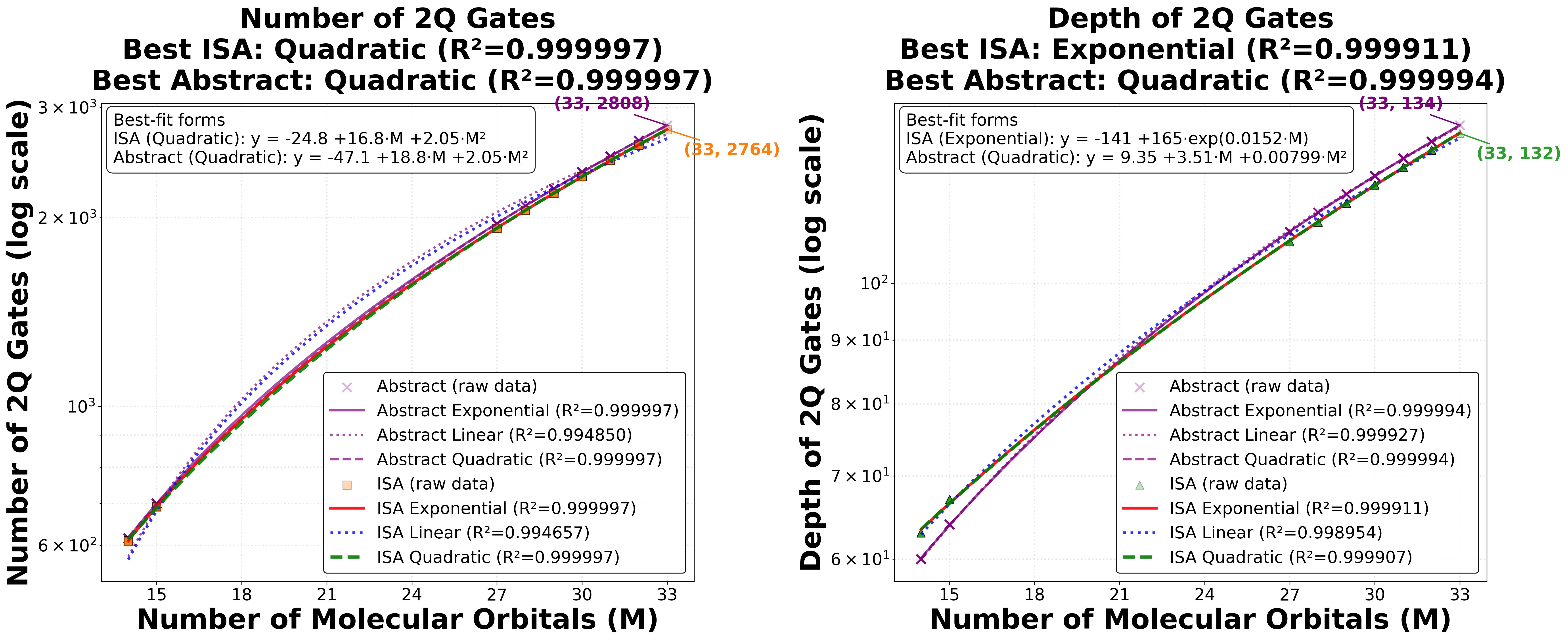}
\caption{\textbf{Size of quantum circuits.} Two-qubit gate count and two-qubit circuit depth of the LUCJ fragment circuits versus orbital count $\norb$, at the abstract and ISA (post-transpilation) levels. The gate count is quadratic in $\norb$ at both levels. Over the present range ($\norb=13$--$33$) the ISA depth is described equally well by quadratic and weakly exponential fits ($\Delta R^{2}\sim \scinot{4}{-6}$); we treat it as polynomial and note that a broader span of fragment sizes is needed to resolve whether asymptotic quadratic, exponential, or other scaling is better supported.}
\label{fig:circuit_parameter_scaling}
\end{figure*}

We sample these candidate configurations from a local unitary cluster Jastrow (LUCJ) ansatz\cite{motta2023bridging} with a single layer ($n_{\mathrm{reps}}=1$), nearest-neighbor $\alpha\alpha$ entangling indices, and $\alpha\beta$ couplings on every fourth orbital when allowed by quantum hardware connectivity. The circuits are implemented through the \texttt{ffsim} library\cite{sung2026ffsim}, with Qiskit\cite{javadi2024quantum} handling transpilation and the IBM \texttt{Sampler} primitive driving the circuit execution on the Heron r3 processor \device{boston}. The LUCJ amplitudes are seeded from the classical CCSD $T_1$ and $T_2$ tensors of the same embedded cluster and then refined with an L-BFGS-B optimizer\cite{lin2025improved}. The count of the two-qubit gates of these circuits scales quadratically with the fragment orbital count $\norb$ (Figure~\ref{fig:circuit_parameter_scaling}). Transpilation onto the limited heavy-hex connectivity of \device{boston} adds SWAP gates, so the hardware (ISA) circuits carry more two-qubit gates and greater depth than the abstract circuits, while preserving the quadratic scaling expected for the fixed-depth ($L=1$) ansatz. The largest fragment treated here ($\norb=33$; cluster~4, fragment~15) maps to a 66-qubit circuit, whose hardware layout is shown in Figure~\ref{fig:largest_fragment_layout}.

To control hardware cost while retaining accuracy in the larger embedded spaces, fragments with orbital count $\norb < 13$ are routed to classical FCI, whereas fragments with $\norb \ge 13$ are dispatched to the QPU. This division of labor reflects the physics of the salt. Electron correlation in this ionic--covalent salt is spatially short-ranged and is therefore well captured on the fragment (atom-plus-bath) scale; the largest active spaces arise at the fluorine-centered fragments, which reach 27--33 orbitals. The corresponding 54--66-qubit Hilbert spaces fall within the window addressable by present-day quantum processors. Within the QPU set, the shot budget is set by an orbital cutoff of 20, with $10^5$ shots for $\norb < 20$ and $10^6$ shots for $\norb \ge 20$.

After identifying a subspace with SQD, accuracy is further improved by extracting the most relevant configurations and augmenting them through single-excitation operators (the subspace-extension stage of Figure~\ref{fig:sqdmethods}). This extended-SQD (ext-SQD) step produces the eigenvalue and reduced density matrices that we report for each QPU-targeted fragment\cite{barison2025quantum,shirakawa2025closed,lin2025improved,walkup2026scaling,kim2024distributed,javadi2024quantum}. The ext-SQD reduced density matrices feed back into the same partitioned-cumulant assembly used for EWF--FCI.

\textbf{Conformational energies of FLiBe clusters.} We benchmark single-point energies of nine cluster conformations of a \ce{Li6Be3F12} cluster, sampled from an ab-initio molecular dynamics (AIMD) trajectory of molten FLiBe against a ladder of first-principles methods: restricted Hartree--Fock (RHF), second-order M\o{}ller--Plesset perturbation theory (MP2), full-system CCSD evaluated in PySCF\cite{sun2018pyscf,sun2020pyscf}, domain-based local-pair natural-orbital DLPNO-CCSD(T)\cite{riplinger2013dlpno} evaluated in ORCA\cite{neese2012orca,neese2022orca}, 
EWF with CCSD (\ewf{CCSD})\cite{nusspickel2022systematic} and FCI (\ewf{FCI}) as fragment solver. The heterogeneous quantum-classical \ewf{FCI+ext-SQD}, diagonalizes the larger fragments (13 or more spatial orbitals) on IBM superconducting quantum processors using ext-SQD\cite{robledo2025chemistry,motta2024subspace,lassqd,shirakawa2025closed,lin2025improved,barison2025quantum,sung2026ffsim,walkup2026scaling,kim2024distributed}.
We report energy differences $\Delta E = E_{R,\mathrm{method}}-E_{R_1,\mathrm{\ewf{FCI}}}$ in kcal/mol using cluster 1 at the EWF-FCI level as reference.

\begin{figure}[htbp]
  \centering
  \includegraphics[width=\linewidth]{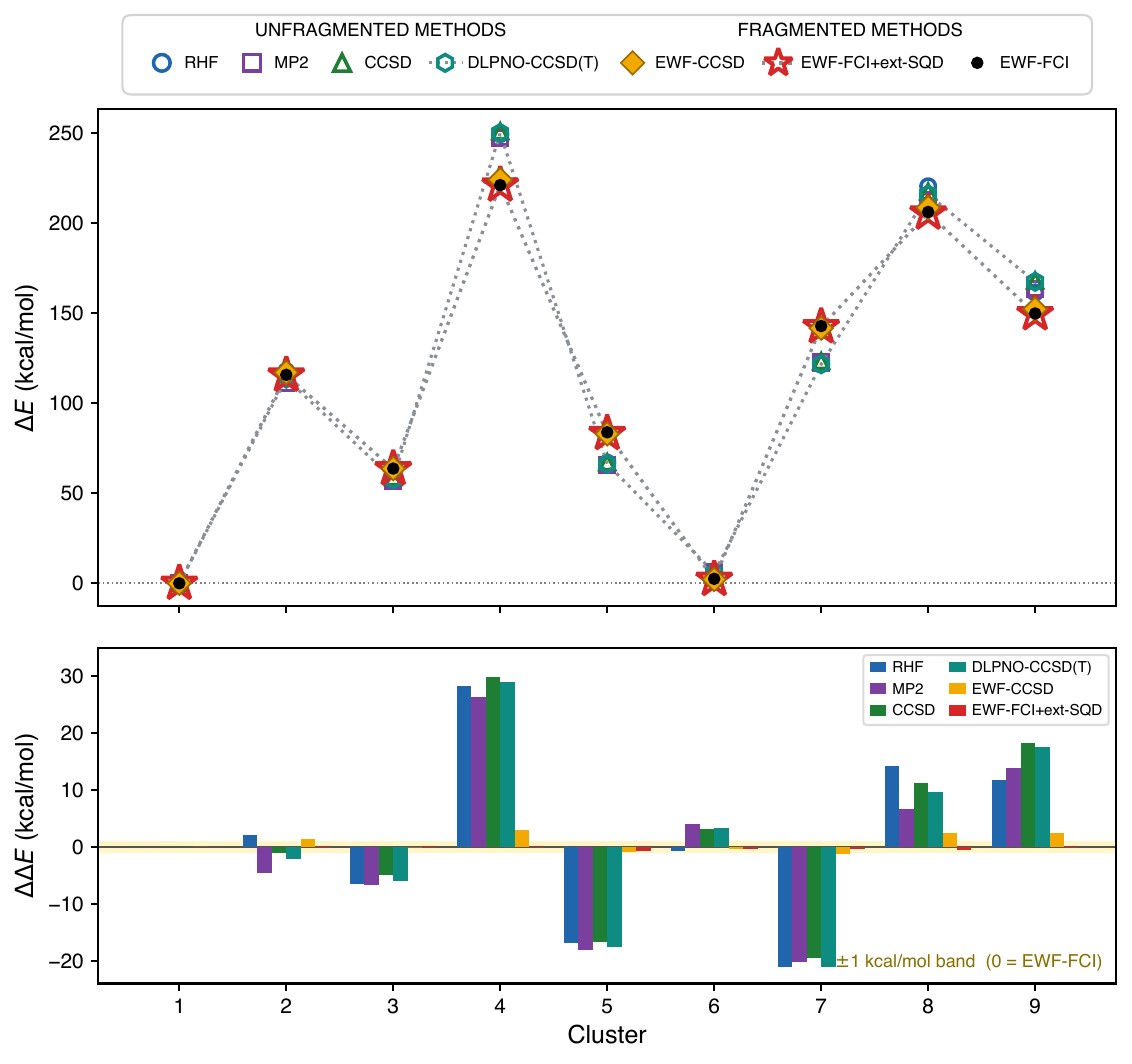}
  \caption{Top panel: relative energy $\Delta E = E_R - E_{R_1}$ (kcal/mol) of nine 21-atom \ce{Li6Be3F12} clusters, each referenced to cluster~1. The four full-molecule (unfragmented) methods (RHF, MP2, DLPNO-CCSD(T), and canonical CCSD) are drawn as open markers and grouped at the left of the legend; the three embedded (fragmented) methods (\ewf{CCSD}, drawn as a filled diamond, \ewf{FCI+ext-SQD}, and \ewf{FCI}) at the right. The quantum-hardware result \ewf{FCI+ext-SQD} is drawn as a large open red star and \ewf{FCI} as a small filled black circle. Dotted lines connect the nine \ewf{FCI+ext-SQD} values and the nine DLPNO-CCSD(T) values across clusters, and the vertical separation between these two trends marks the embedding gap between the best full-molecule and embedded references. Bottom panel: deviation of each method from the \ewf{FCI} reference, $\Delta\Delta E = \Delta E_{\mathrm{method}} - \Delta E_{\mathrm{\ewf{FCI}}}$, drawn as grouped bars per cluster in the legend colours (unfragmented methods together with the fragmented \ewf{CCSD} and \ewf{FCI+ext-SQD}); the $\pm 1$~kcal/mol accuracy window is shaded in gold and the zero line is the \ewf{FCI} reference. The unfragmented bars deviate by up to ${\sim}30$~kcal/mol, whereas the fragmented bars remain within a few kcal/mol of \ewf{FCI}, the ext-SQD deviations lying inside the gold band.}
  \label{fig:dE9confs}
\end{figure}

All nine clusters are single-reference by the $T_1$ diagnostic ($T_1 = 0.014$--$0.016$), so the comparison below is not affected by multireference character. We first compare the three EWF solvers. \ewf{CCSD}, \ewf{FCI}, and \ewf{FCI+ext-SQD} share an identical embedding and differ only in the fragment solver (Methods). Within this family, the ext-SQD solver agrees with FCI to within 0.7 kcal/mol in $\Delta E$, at a mean absolute deviation of 0.3 kcal/mol (Figure~\ref{fig:dE9confs}). Classical \ewf{CCSD} reproduces the same reference to within 3.0 kcal/mol. The quantum-generated energies lie $+2.1$ to $+2.9$ kcal/mol above FCI in absolute terms, which we ascribe to residual device noise and limited determinant sampling/diagonalization. This offset is nearly constant across clusters, so it cancels in relative energies. On a charged, ionic, single-reference system, with fragments of 13-33 orbitals, ext-SQD therefore solves fragment ground-states with accuracy comparable to FCI. To our knowledge this is the first such demonstration for an ionic, charged-species system, and a starting point to extend \ewf{SQD} from neutral protein active sites\cite{shajan2025molecular} to ionic--covalent chemistry and beyond.

The conventional full-system methods behave differently. RHF, MP2, CCSD, and DLPNO-CCSD(T) agree closely with one another, but all four sit 11--12 kcal/mol from EWF-based calculation, and nearly 30 kcal/mol at cluster 4. Even DLPNO-CCSD(T), normally a near-quantitative benchmark, tracks the other three rather than the embedded result. The discrepancy between the two families stems from the fragmentation approximation, not the quality of the fragment solutions. Indeed, the three embedded methods share a common fragmentation scheme and agree with each other, while the four full-system methods have no fragmentation and agree with each other. The only step that distinguishes the two families is the fragment construction, defined by a bath truncation set to $\eta = 10^{-5}$. The discrepancy therefore quantifies the impact of the fragmentation approximation, which is the dominant source of bias in the present study (more than an order of magnitude larger than the 0.7 kcal/mol solver error), and requires lowering $\eta$ to be mitigated.

The ordering of these two errors is what matters for the free-energy application. Molten-salt observables are Boltzmann averages over configuration ensembles of exactly the kind sampled here, and an error cancels in such an average only when it is uniform across configurations. Deviations between embedded and full-system energies are not uniform. They vary by tens of kcal/mol across AIMD snapshots and change sign (bottom panel of Figure~\ref{fig:dE9confs}). At $k_B T \approx 1.8$~kcal/mol near the blanket operating temperature of 900~K, the 0.7 kcal/mol solver error shifts an equilibrium constant by less than a factor of 1.5, whereas a configuration-dependent fragmentation error, which reaches 11--30 kcal/mol across these snapshots and changes sign, dominates any computed free energy. The quantum solver is, therefore, already accurate enough. The task that remains is to converge the embedding, which we will address in future work.

\textbf{Tritium binding energies.} The conformational energies of the previous section tested the workflow on the neutral salt. We now turn to the quantity that motivates the study: the energy with which tritium binds in FLiBe. We define the tritium binding energy as $E_{\mathrm{bind}} = E_{\mathrm{FLiBeTF}} - E_{\mathrm{FLiBeF}^-}$, the difference between the 23-atom tritiated cluster and the 22-atom fluoride-terminated anion, with the tritium removed as a bare \ce{T+} nucleus. $E_{\mathrm{bind}}$ is a ground-state electronic binding energy evaluated at a single fixed AIMD geometry, not itself a free energy; the free energy follows only from its Boltzmann average over configurations (Supporting Information). This is the insertion energy $\varepsilon = U_{N+1} - U_{N}$ that enters the quasi-chemical free-energy correction~\cite{shi2022deep} introduced above. Its ensemble average over salt configurations sets the excess chemical potential of tritium, and with it the thermodynamic cost of breeding and recovering tritium in the blanket. An accurate binding energy is therefore a necessary input to any predictive treatment of tritium speciation.

\begin{figure*}[htbp]
  \centering
  \includegraphics[width=0.98\linewidth]{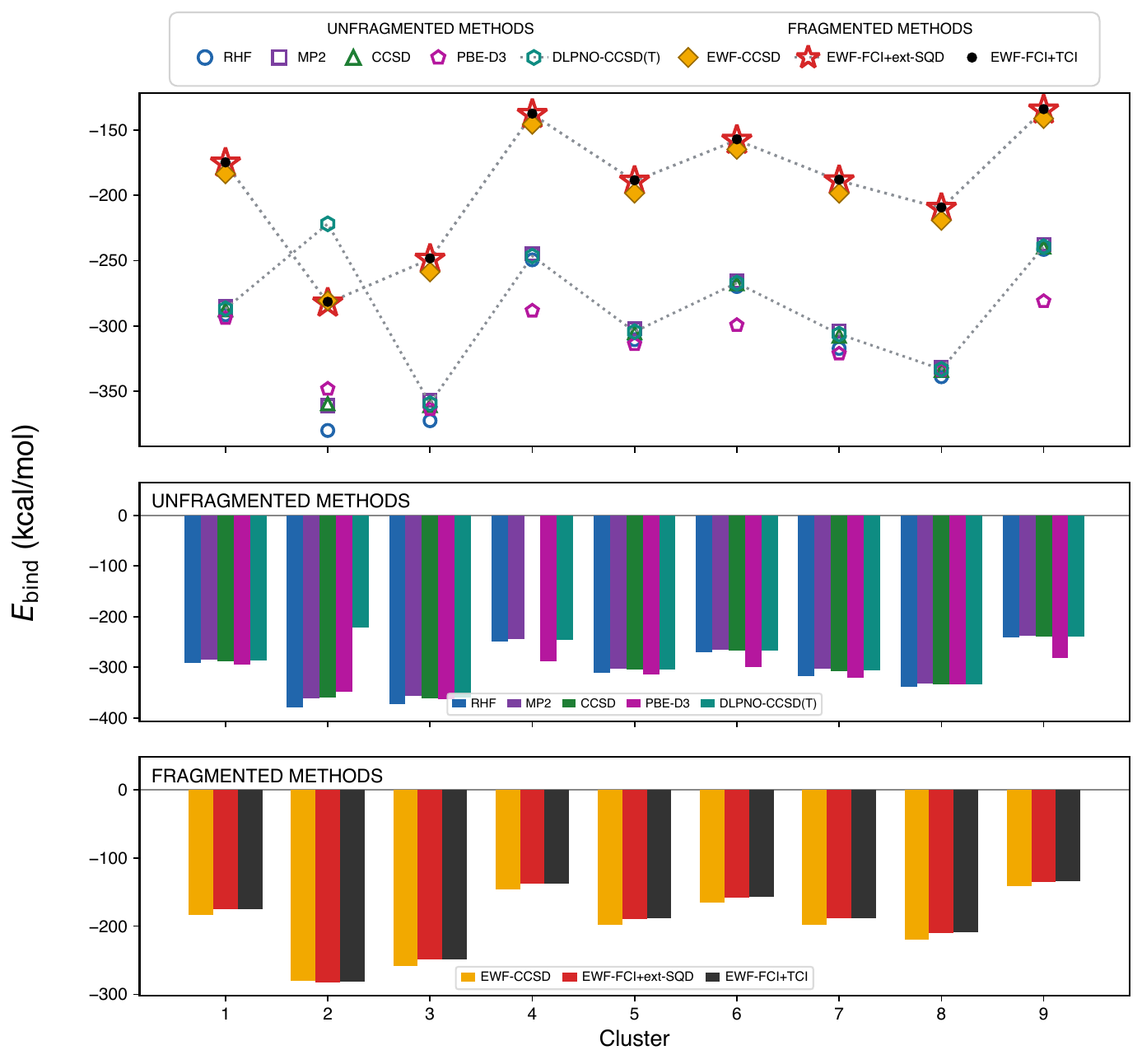}
  \caption{\textbf{Tritium binding energy to FLiBe across nine clusters at eight levels of theory.} The tritium binding energy $E_{\mathrm{bind}} = E_{\mathrm{FLiBeTF}} - E_{\mathrm{FLiBeF}^-}$ (kcal/mol) is the total-energy difference between the 23-atom neutral \ce{Li6Be3F13T} (FLiBeTF) and the 22-atom anionic \ce{[Li6Be3F13]-} (FLiBeF$^-$), with the tritium removed as a bare \ce{T+} nucleus of zero electronic energy. Top panel: markers report, per cluster, RHF, MP2, canonical CCSD, DLPNO-CCSD(T), PBE-D3 (the dispersion-corrected functional used to generate the AIMD/MLFF configurations; Supporting Information), EWF-CCSD,EWF-FCI+ext-SQD, andEWF-FCI+TCI taken as the embedded reference; the five full-molecule (unfragmented) methods are drawn as open markers and grouped at the left of the legend, the three embedded (fragmented) methods at the right.EWF-FCI+ext-SQD is drawn as a large open red star, and dotted lines connect its nine values and the nine DLPNO-CCSD(T) values; the vertical gap between these two trends is the embedding offset discussed in the text. The middle and bottom panels recast the same binding energies as grouped bars in the legend colours, separated by family: the five full-molecule (unfragmented) methods (middle) and the three embedded (fragmented) methods (bottom), sharing the single $E_{\mathrm{bind}}$ axis at the left. Canonical CCSD is omitted at cluster~4, where its FLiBeTF SCF did not converge, and DLPNO-CCSD(T) is anomalously weakly bound at cluster~2. Numerical values are listed in Table~S3.}
  \label{fig:binding}
\end{figure*}

Two features stand out (Figure~\ref{fig:binding}; Table~S3). First, the quantum-hardware solver reproduces its classical embedded parent even for this charged, cross-system difference.EWF-FCI+ext-SQD matchesEWF-FCI+TCI to a mean absolute deviation of 0.7~kcal/mol, with a maximum of 0.9~kcal/mol across all nine conformations. Here TCI denotes TCI-8, a highly accurate truncated configuration-interaction solver used as the classical reference for the five largest FLiBeTF fragments, whose 30--31 orbital active spaces place exact FCI beyond present resources; TCI-8 reproduces FCI to within a microhartree (Supporting Information).The ext-SQD path therefore solves the embedded binding problem at the accuracy of  the truncated-CI reference it replaces, whereas classical EWF-CCSD departs from that reference by as much as 10~kcal/mol. Second, every method agrees that the tritium ion is strongly bound, with binding energies spanning $-134$ to $-380$~kcal/mol. This range is consistent with the tightly coordinated ionic tritium seen in finite-temperature AIMD of FLiBe melts\cite{lam2021impact,wang2022first}. The absolute values nonetheless separate into two bands: the three embedded methods fall between $-134$ and $-282$~kcal/mol, and the five full-molecule methods between $-222$ and $-380$~kcal/mol, an offset of about 110~kcal/mol. As in the conformational energies, this offset is a signature of the embedding rather than of any solver, and here it does not cancel. In the relative energies of Figure~\ref{fig:dE9confs} the embedding error drops out, but the binding energy is a difference between two \emph{separately embedded}, oppositely charged systems, so the error survives and sets the dominant spread between methods. That ext-SQD tracks FCI to within kcal/mol accuracy in fragment solution under this harder test is an encouraging indication of a favorable accuracy-cost tradeoff. The embedding is the limiting error source, and reducing requires improved fragment construction.

\textbf{Outlook}

This study provides an initial assessment of heterogeneous quantum-classical computing for investigating the interaction between tritium and molten salts, exemplified by FLiBe. Our analysis identifies fragment size as an important source of error at the current scale, and shows that ext-SQD can provide kcal/mol-accurate solutions of individual fragments with up to ($\norb \leq 33$) orbitals. 
While these results are encouraging, they also indicate limitations and directions for the development of higher-accuracy heterogeneous quantum-classical simulations for molten-salt chemistry.

Achieving the long-term goal of predictive free-energy simulations of tritium in realistic molten-salt environments, which requires $\approx k_B T$ accuracy over simulation cells containing in excess of 100 atoms, will require advances in several areas.
A key direction is increasing the size of clusters studied towards the thermodynamic limit of large system size, which is necessary for any calculation representative of a liquid.
The increase in system size has to be accompanied by the use of chemically realistic (e.g. correlation consistent) basis sets, and a more extensive (and possibly improved) sampling of cluster configurations.
At the level of EWF calculations, key directions are chemically informed fragment construction, larger fragments, and more efficient quantum-classical methods to solve the fragment Schr\"{o}dinger equation. Increasing fragment size should be based on a chemically aware strategy, more sophisticated than decreasing the fragmentation threshold $\eta$, to account for chemical bonding and electronic correlations while controlling computational cost. Indeed, as active-space size grows, maintaining high-accuracy ext-SQD results requires an increasing number of electronic configurations, which increases pressure on classical matrix diagonalization and quantum sampling in the presence of more intense hardware noise.

From the theoretical viewpoint, an important direction of future research will be rationalizing how the bath truncation translates into methodological biases: fragments may be spatially localized, losing long-range interactions that lead to dispersion interactions, if they could be concentrated in the valence space, losing contributions from polarized and diffused basis functions. This assessment will allow to gauge the effectiveness of lowering $\eta$ in improving EWF accuracy, or guide the potential development of new methods to treat e.g. long-range interactions. Furthermore, whereas the cluster conformations reported here feature single-reference electronic states, we observed that other conformations have multireference character, and their investigation will be the topic of future work.

A particularly promising direction in terms of improving the accuracy-cost tradeoff of our heterogeneous quantum-classical workflow is the integration of AI elements. Machine-learned interatomic potentials can accelerate configurational sampling, while regression and classification AI methods may enable automated construction of molecular orbitals, chemically meaningful fragments, optimized quantum circuits, and compact yet accurate representations of electronic ground-state wavefunctions. The latter goal may be accomplished removing ``deadwood'' configurations without the need of iterative procedures, and adding ``livewood'' configurations not sampled by the quantum computer.
The high-accuracy quantum-classical calculations developed in this work provide a natural source of training and validation data for such AI models.

More broadly, this work contributes to an integrated AI-driven discovery cycle aimed at optimizing molten salts for fusion energy applications, within a broader DOE Genesis activity. In this context, high-fidelity quantum-HPC calculations of tritium chemistry inform surrogate models, which in turn guide the exploration of molten-salt compositions and operating conditions.

Finally, the computational challenges encountered in accurate molten-salt free-energy calculations (particularly the need to efficiently sample atomic configurations and deliver accurate correlated electronic wavefunctions) are found across other areas of chemistry, including catalysis and biochemistry. Improving the accuracy-cost tradeoff of heterogeneous quantum-classical methods with molten salts as a target system can provide tangible benefit transferrable to other fields of quantum chemistry.

The AIMD/MLFF configurational sampling was performed on Perlmutter (NERSC) and Frontier (OLCF), the reference FCI calculations on Frontier, and all mean-field, EWF, full-molecule benchmark, and SQD/ext-SQD post-processing on the Michigan State University HPCC and the Cleveland Clinic Foundation HPC, while quantum sampling ran on the IBM Heron~r3 processor \device{boston}; a full QPU, GPU, and CPU breakdown by system and method is given in the Supporting Information.

\textbf{Acknowledgments}

V.S.B and T.J.P.S. are supported by the Molten Salts in Extreme Environments (MSEE) Energy Frontier Research Center, funded by the U.S. Department of Energy (DOE) Office of Science, Office of Basic Energy Sciences under the DOE contract DE-AC05-00OR22725 with Oak Ridge National Laboratory (ORNL). This research used resources of the Oak Ridge Leadership Computing Facility at ORNL, which is supported by the Advanced Scientific Computing Research programs in the Office of Science of the U.S. Department of Energy under Contract No. DE-AC05-00OR22725. This research also used resources of the National Energy Research Scientific Computing Center (NERSC), a Department of Energy User Facility using NERSC award BES-ERCAP-0035934. The authors gratefully acknowledge financial support from the National Science Foundation (NSF) through CSSI Frameworks Grant OAC-2209717 and from the National Institutes of Health (Grant Numbers GM130641). The authors are grateful to the high-performance computer center (iCER HPCC) at Michigan State University and the high-performance computer center at Cleveland Clinic Foundation. We thank George H. Booth for helpful discussion on the EWF method which was essential for scaling of the simulations within this method. 

\clearpage
\setcounter{secnumdepth}{3}
\setcounter{section}{0}
\setcounter{figure}{0}
\setcounter{table}{0}
\setcounter{equation}{0}
\renewcommand{\thesection}{S\arabic{section}}
\renewcommand{\thefigure}{S\arabic{figure}}
\renewcommand{\thetable}{S\arabic{table}}
\renewcommand{\theequation}{S\arabic{equation}}
\renewcommand{\thepage}{S\arabic{page}}
{\centering\noindent\Large\textbf{Supporting Information}\par}
\vspace{1em}
\section{Benchmark Systems and Clusters Geometries}

\begin{figure}[h]
  \centering
  \includegraphics[width=\linewidth]{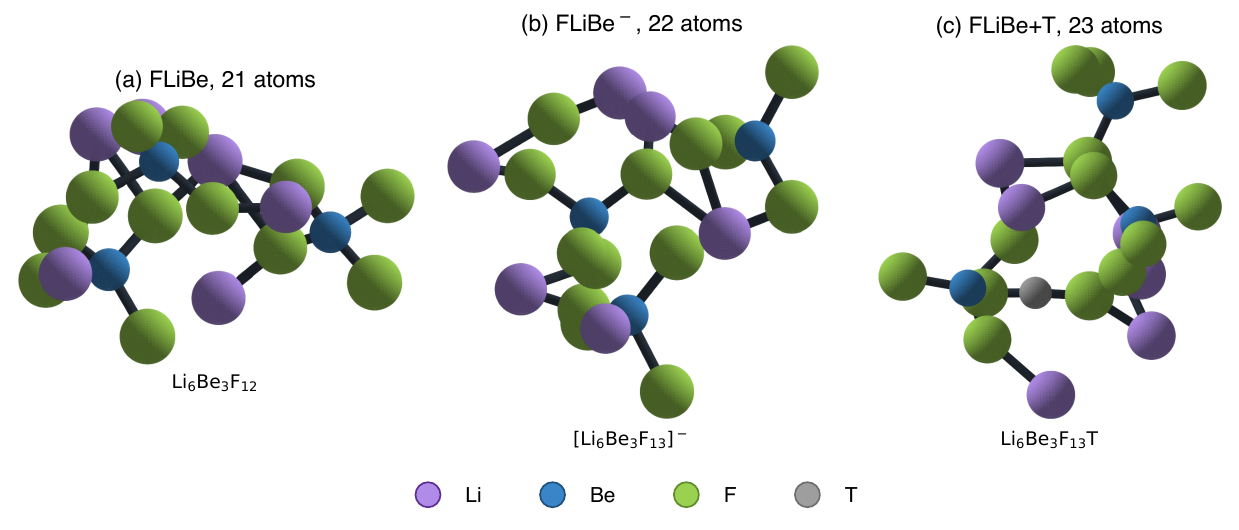}
  \caption{Representative ball-and-stick cluster geometries for the systems benchmarked in this work. (a) Canonical 21-atom FLiBe (\ce{Li6Be3F12}, System 1). (b) The 22-atom deprotonated reference \ce{[Li6Be3F13]^-} (FLiBe$^-$, charge $-1$) and (c) the 23-atom tritium-bound form \ce{Li6Be3F13T} (FLiBe$+$T, charge $0$) used to define the tritium binding energy of System 2. Each beryllium is tetrahedrally coordinated by four fluorines (\ce{BeF4}), and in (c) the tritium bridges two fluorines in an F--T--F motif. Color code: Li (violet), Be (steel blue), F (lime), T (gray).}
  \label{fig:systems}
\end{figure}

Three molecular systems are benchmarked in this work (Figure~\ref{fig:systems}): the neutral 21-atom FLiBe cluster \ce{Li6Be3F12} (System~1), and the tritium-binding pair of System~2, namely the 23-atom \ce{Li6Be3F13T} (FLiBeTF, charge $0$) and the 22-atom \ce{[Li6Be3F13]-} (FLiBeF$^-$, charge $-1$). Each system comprises nine clusters, fragmented into one embedded cluster per atom, giving 189 (FLiBe), 207 (FLiBeTF), and 198 (FLiBeF$^-$) embedded fragments in total. The embedded-cluster size distributions of the two System-2 species are compared in Figure~\ref{fig:fragsize}(b).

The cluster geometries were extracted from the finite-temperature AIMD/MLFF trajectories of bulk molten FLiBe described in Section~\ref{sec:clustergen}, sampled at well-separated timesteps to capture distinct local coordination environments of the ionic Li--F and partially covalent Be--F networks. The benchmark systems contain 21 (\ce{Li6Be3F12}), 22 (\ce{[Li6Be3F13]-}, FLiBeF$^-$), and 23 (\ce{Li6Be3F13T}, FLiBeTF) atoms, and every cluster considered here is treated as a closed-shell singlet in the gas phase. The 22-atom anion has the same atomic coordinates as the 23-atom FLiBeTF cluster with the tritium removed.

\section{Free energy perturbations}

{\bf Free Energy Correction.}  Here we describe the perturbation approach which outlines the procedure for computing the free energy corrections to the tritium binding free energies.  This method will be used in future studies of tritium in FLiBe.  

The free energy change for the transition from an approximate DFT result to the value produced using data from the quantum computer generated energies is
\begin{equation}
\label{eq:fe}
\Delta \mu^{\mathrm{ex}} = -kT \ln{\langle \exp({-\Delta \varepsilon / kT)} \rangle_{\mathrm DFT}}
\end{equation}
Here $\Delta \mu^{\mathrm{ex}}$ is the excess chemical potential (free energy) change going from the DFT prediction to the quantum computing generated prediction, $k$ is Boltzmann's constant, $T$ is the temperature, and $\Delta \varepsilon$ is the difference of the ion binding energies computed at the quantum computing and DFT levels ($\Delta \varepsilon = \varepsilon_{\mathrm{QC}} - \varepsilon_{\mathrm{DFT}}$ where $\varepsilon$ is the binding energy, or the difference in total ground state energies $\varepsilon = U_{N+1} - U_N$ upon adding the ion to the $N$-ion cluster).  The DFT subscript on the statistical average implies that the configurations are sampled at the DFT level.  The formula is exact for the free energy correction.  In Ref.~\citenum{shi2022deep} the statistical average $\langle \ldots \rangle$ was calculated using 400 well-separated (independent) configurations during the DFT simulation.

\section{AIMD/MLFF cluster generation}\label{sec:clustergen}

{\bf Equilibrium atomic clusters.} The atomic clusters studied in this work were extracted from randomly selected configurations sampled from molecular dynamics (MD) trajectories of larger periodic systems. Trajectories were generated using the Vienna \textit{Ab initio} Simulation Package\cite{Kresse1996} (VASP) in the isothermal-isobaric ensemble (NpT) employing a Langevin thermostat\cite{Allen1991} with a friction coefficient of 10 ps$^{-1}$, while pressure was maintained at 1 bar using the Parrinello–Rahman algorithm\cite{Parrinello1980,Parrinello1981} with a friction coefficient of 5 ps$^{-1}$ and a fictions lattice mass of 1000 amu applied to all cell degrees of freedom. The integration timestep was 2 fs. To efficiently sample the configurational space while retaining near-DFT accuracy, trajectories were continuously generated over a temperature range of 900–1300 K using an active learning framework that coupled \textit{ab initio} molecular dynamics (AIMD) with on-the-fly training of a kernel-based machine learning force field (MLFF).\cite{Jinnouchi2019} Starting from a  pre-equilibrated configuration, the MLFF was iteratively trained on density functional theory (DFT) energies, forces, and stresses from configurations exhibiting the highest uncertainty, as identified through a Bayesian-guided active learning framework.\cite{Jinnouchi2019,Jinnouchi2020} For the tritium-containing system, the training dataset was further augmented with metadynamics simulations performed in the canonical ensemble (NVT) at 900 K using a Nosé–Hoover thermostat\cite{Martyna1992,Evans1985} with a mass parameter (SMASS) of 0.32, where the collective variable was defined as the coordination number between tritium and fluoride ions to enhance the sampling of distinct coordination environments. The corresponding DFT reference calculations were subsequently performed with the CP2K package\cite{Kuhne2020} using the Perdew–Burke–Ernzerhof (PBE) exchange-correlation functional\cite{Perdew1996,Perdew1996b,Zhang1998} and Grimme's D3 dispersion correction,\cite{Grimme2010} with Goedecker–Teter–Hutter (GTH) pseudopotentials\cite{Goedecker1996,Hartwigsen1998} and TZV2P-MOLOPT Gaussian basis sets. The cutoff of the finest multigrid level was set to 2000 Ry, and the plane wave cutoff of a reference grid for mapping the Gaussian basis functions was set to 100 Ry. 

\begin{figure}[ht!]
  \centering
  \includegraphics[width=0.6\linewidth]{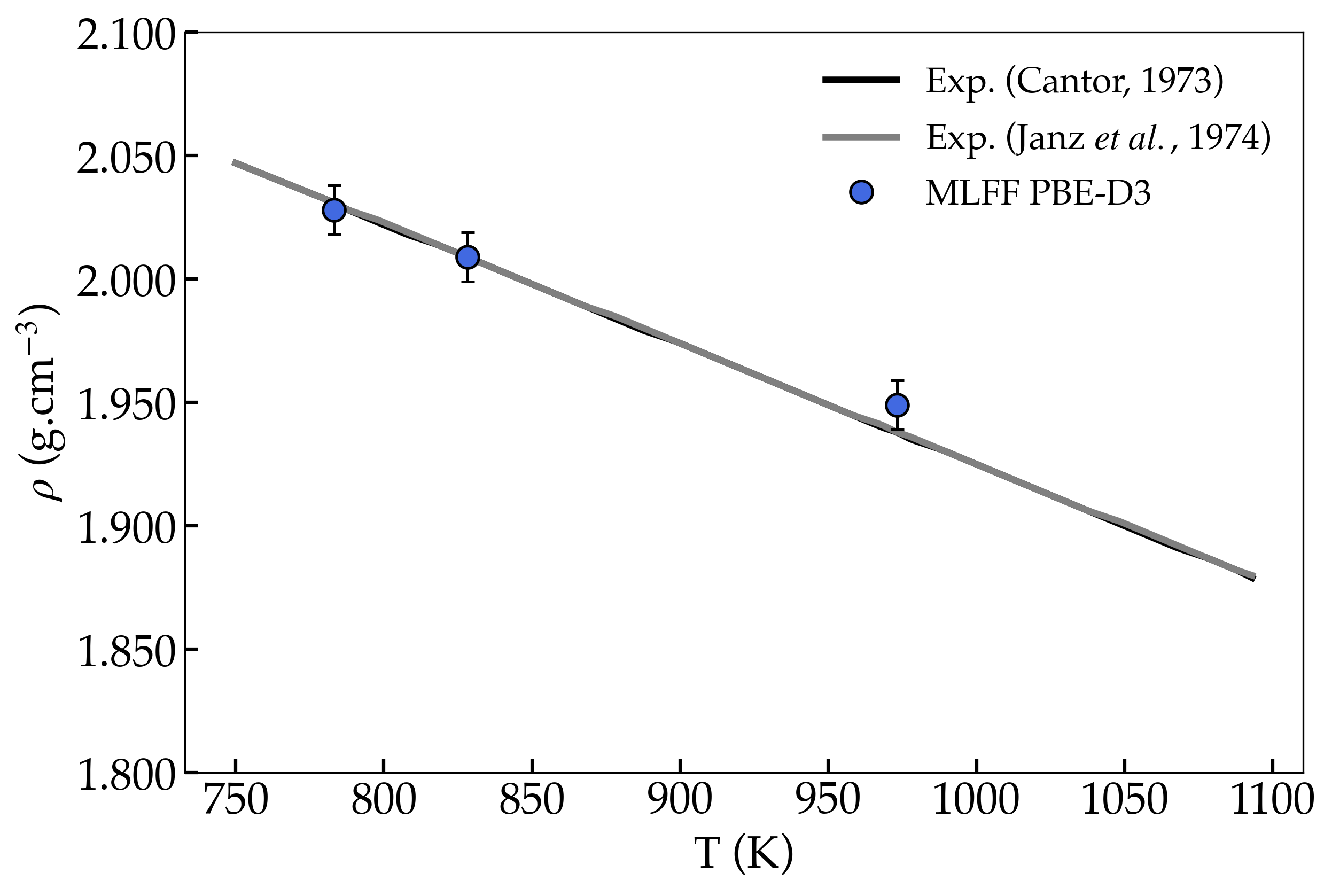}
  \caption{Temperature dependence of the density of pure FLiBe obtained from MLFF PBE-D3 simulations compared with experimental data,\cite{Cantor1973,Janz1974} demonstrating excellent agreement across the investigated temperature range. Error bars denote one standard deviation.}
  \label{fig:S1}
\end{figure}

\begin{figure}[ht!]
  \centering
  \includegraphics[width=0.45\linewidth]{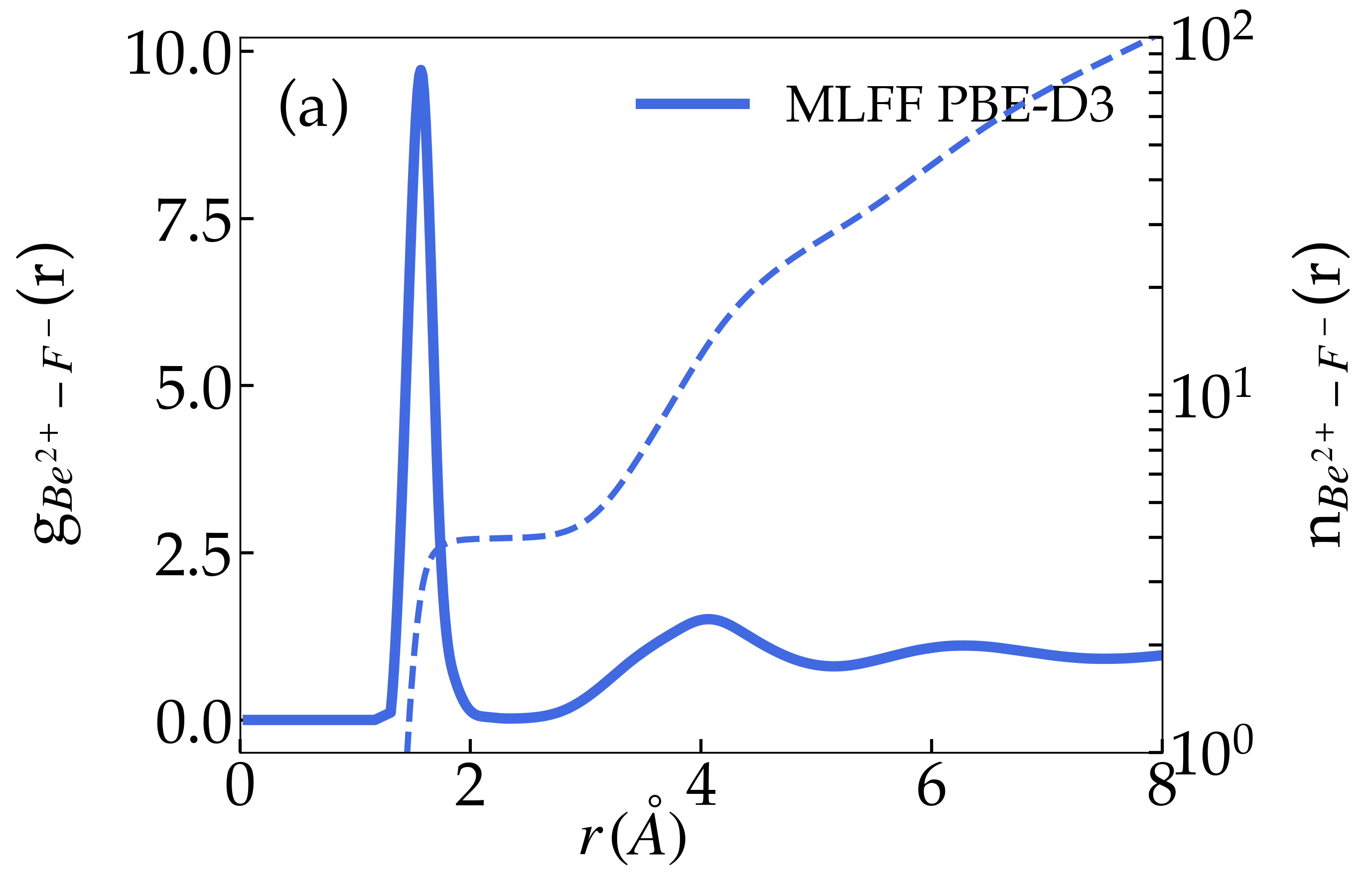}
  \includegraphics[width=0.45\linewidth]{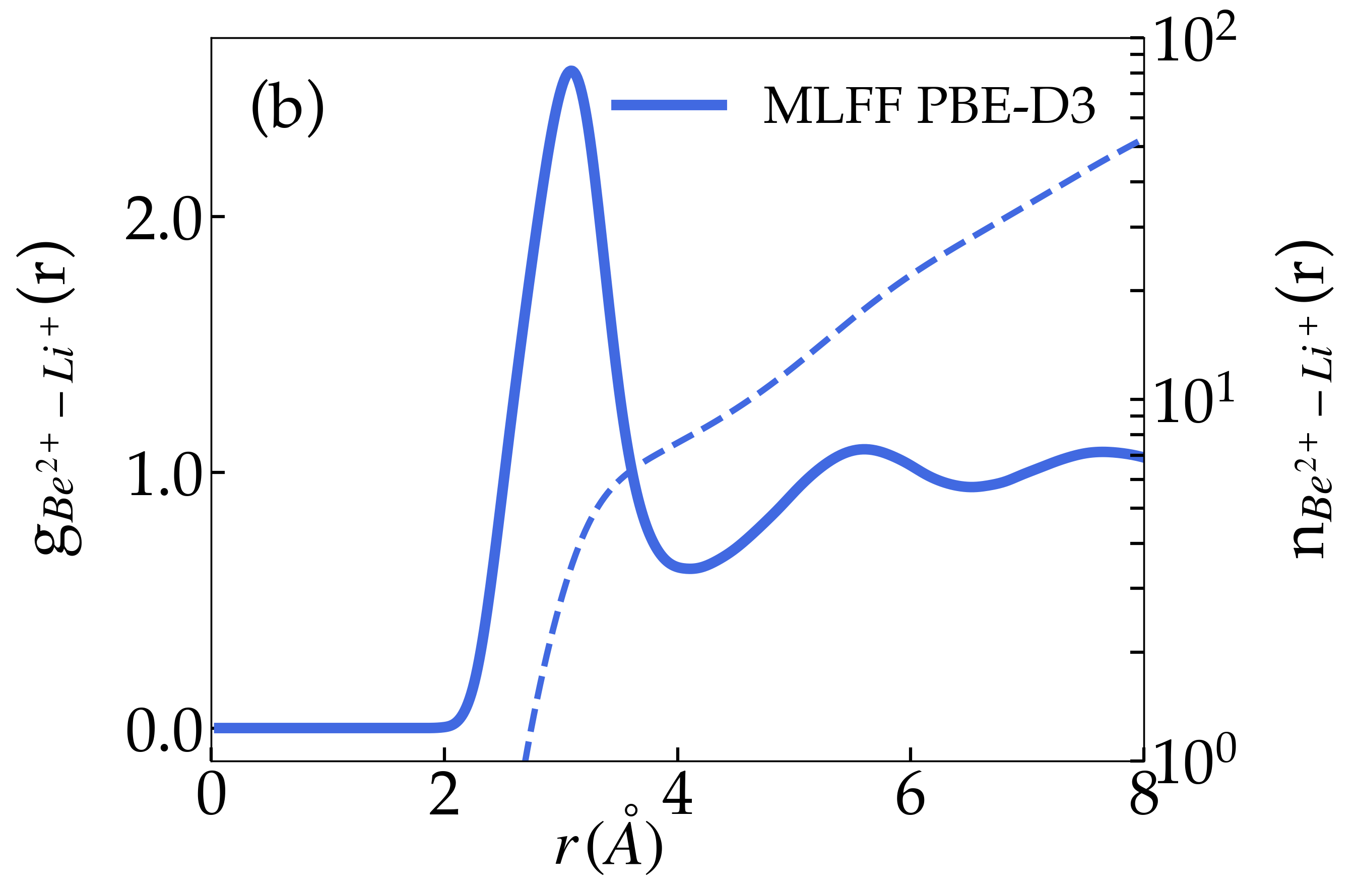}
  \includegraphics[width=0.45\linewidth]{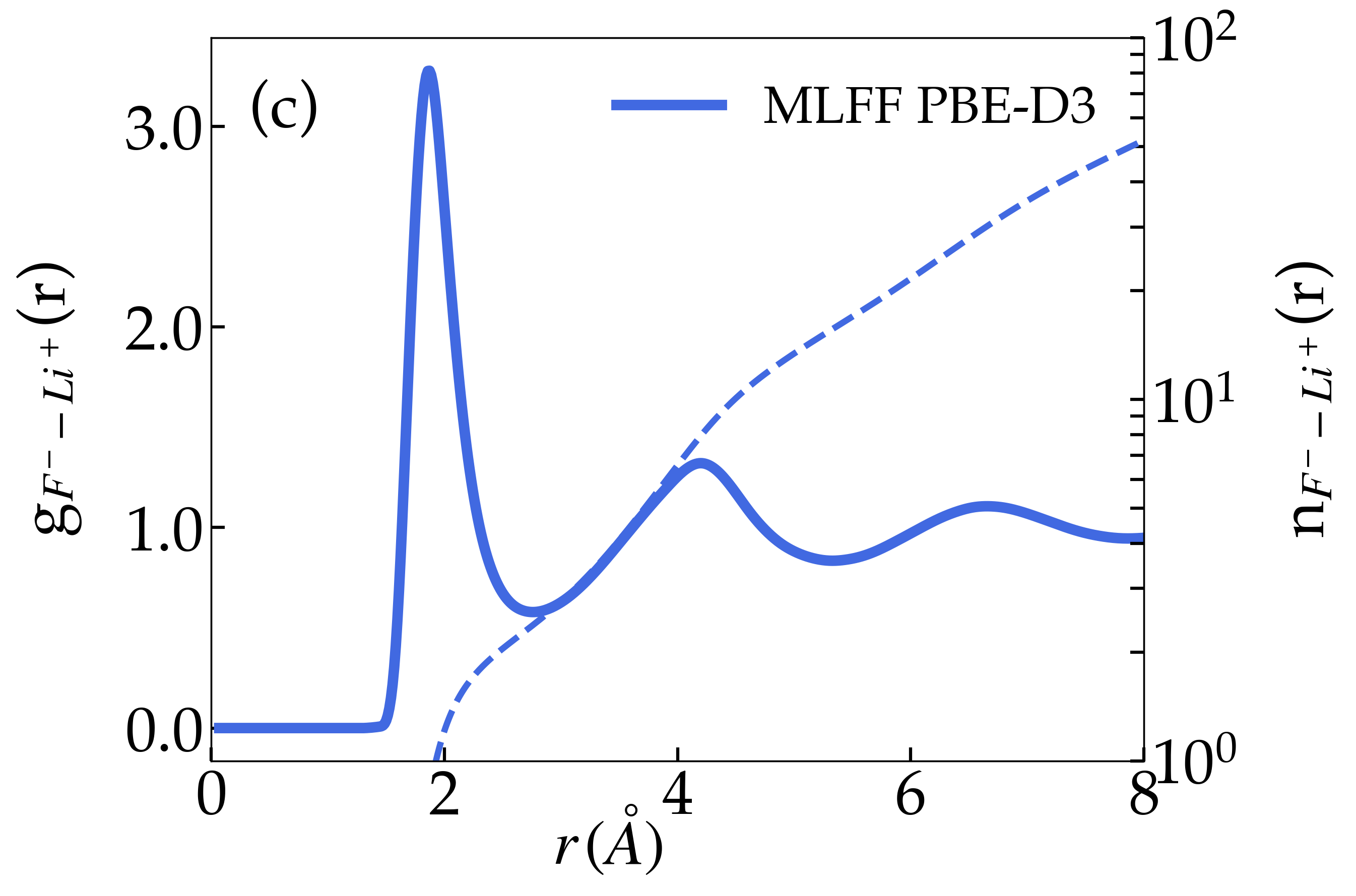}
  \caption{Radial distribution functions, $g(r)$ (solid lines, left axis), and cumulative coordination numbers, $n(r)$ (dashed lines, right axis), for the (a) Be$^{2+}$--F$^{-}$, (b) Be$^{2+}$--Li$^{+}$, and (c) F$^{-}$--Li$^{+}$ pairs in pure FLiBe at 783.15~K, obtained from molecular dynamics simulations using the machine learning force field (MLFF) at the PBE-D3 level of theory. The Be$^{2+}$--F$^{-}$ coordination number approaches 4 at the first minimum, demonstrating the predominance of tetrahedrally coordinated Be$^{2+}$ ions.}
  \label{fig:S2}
\end{figure}

{\bf Cluster extraction and construction.} The pure FLiBe clusters were extracted from equilibrated MLFF trajectories at 783.15 K, whereas the tritium-containing clusters were extracted from trajectories generated during the final stages of the active learning procedure. Results show that the production simulations accurately reproduced the experimental density of pure FLiBe over a broad temperature range (Figure \ref{fig:S1}),\cite{Cantor1973,Janz1974} while radial distribution functions confirmed the expected BeF$_4^-$ coordination (Figure \ref{fig:S2}).
Clusters were generated by randomly selecting configurations from the trajectory and then choosing either a beryllium atom (for FLiBe clusters) or the tritium atom (for tritium-containing clusters) as the initial point for a depth-first search (DFS) algorithm. The search first visited the local neighbors of the target atom and validated their identities against a predefined set of allowed ions before recursively progressing to neighbors-of-neighbors. Neighboring atoms were recursively retained through a local search designed to enforce FLiBe stoichiometry (1 BeF$_2$ : 2 LiF) and preserve the connectivity of the first coordination shell around Be$^{2+}$, which is composed on average of four F$^-$ ions. The cluster size was capped at 21 atoms for pure FLiBe and 23 atoms for the $^{3}$H$^{+}$-containing clusters, where one additional fluoride ion was included to maintain overall charge neutrality.

\begin{figure}[h!]
  \centering
  \includegraphics[width=0.85\linewidth]{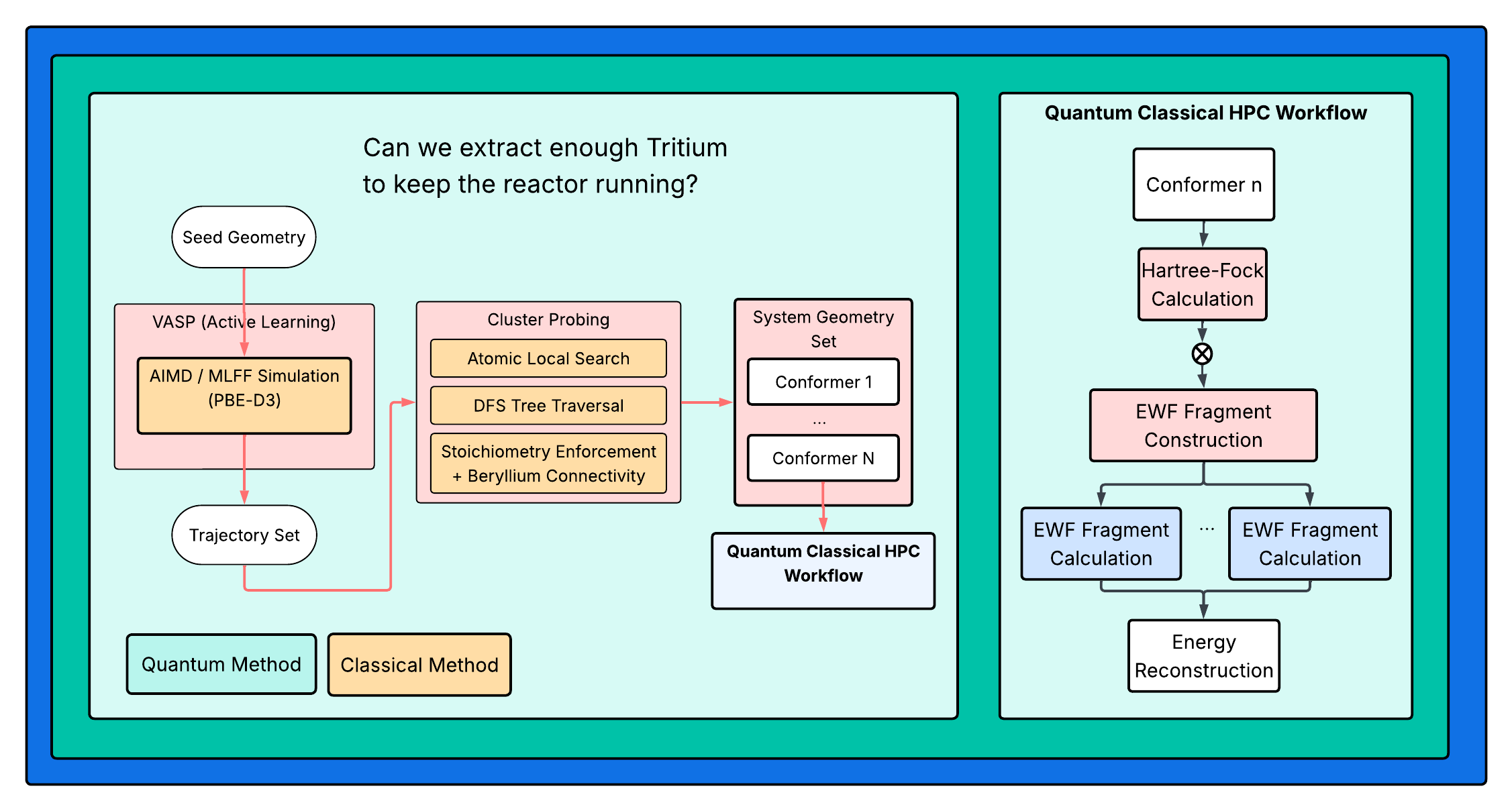}
  \caption{Cluster Probing and Quantum Classical HPC Workflow: Schematic overview of the workflow used to generate the cluster from bulk-phase FLiBe simulations. Active learning-driven AIMD/MLFF simulations at the PBE-D3 level generate representative trajectories, from which random configurations are selected for cluster probing. A depth-first search (DFS) tree traversal algorithm performs an atomic local search that recursively retains neighboring atoms while enforcing FLiBe stoichiometry, charge neutrality, and preservation of the first coordination shell connectivity around Be$^{2+}$ ions. The resulting cluster ensemble is then processed through a Quantum Classical HPC workflow involving Hartree–Fock calculations, embedded wavefunction fragment construction, parallel fragment calculations, and energy reconstruction.}
  \label{fig:ClusterProbing}
\end{figure}

\section{Classical Pre-processing}

\subsection{Mean-field calculations}

All mean-field calculations were performed at the restricted Hartree--Fock (RHF) level in the 6-31+G(d) (hereafter 6-31+G*) Pople basis set. This basis yields 378 atomic orbitals for the neutral \ce{Li6Be3F12} cluster, 396 for the \ce{[Li6Be3F13]-} 22-atom anion, and 398 for the 23-atom \ce{Li6Be3F13T} tritium-binding system. No relativistic correction was applied at the mean-field stage. PySCF~\cite{sun2018pyscf,sun2020pyscf} was used as the SCF backend, called through the Vayesta embedded-wavefunction (EWF) driver~\cite{nusspickel2022systematic,nusspickel2023effective}. The converged RHF density matrix provides the reference one-particle density matrix from which the localized fragments and their entanglement baths are constructed. All mean-field calculations were carried out on the Michigan State University (MSU) High-Performance Computing Center (HPCC). The classical benchmark methods reported in the main text---RHF, MP2, and canonical CCSD in PySCF~\cite{sun2018pyscf,sun2020pyscf} and DLPNO-CCSD(T) in ORCA~\cite{neese2012orca,neese2022orca}---were likewise executed on the MSU HPCC.

\subsection{EWF fragmentation and bath construction}

Because FLiBe combines strongly ionic Li--F bonding with partially covalent Be--F bonding, a chemically faithful localization of the occupied space is essential before fragmentation. We adopt intrinsic atomic orbitals (IAOs)~\cite{knizia2013intrinsic} as the localized one-particle basis. IAOs span the occupied RHF manifold exactly while remaining anchored to atom-centered free-atom reference orbitals. This yields well-defined atomic populations for the formal \ce{Li+}, \ce{Be^2+}, and \ce{F-} ionic charges and avoids the orbital-tail ambiguities of Boys or Pipek--Mezey localizations in this strongly heteropolar system.

In the IAO basis we partition the molecule into one fragment per atom, giving 21 fragments per neutral cluster (22 fragments for the \ce{[Li6Be3F13]-} anion and 23 for the tritium-binding system). For each fragment, the entanglement bath is constructed from a Schmidt decomposition of the RHF one-particle density matrix, in the spirit of density-matrix embedding theory. The Schmidt bath is then expanded at the MP2 level: the MP2 one- and two-body density matrices are diagonalized in the bath-virtual and bath-occupied subspaces to produce bath natural orbitals (BNOs). The BNO occupation numbers measure the correlation-driven coupling between the fragment and its environment. Virtual BNOs with occupation greater than $\eta$ and occupied BNOs with occupation less than $2-\eta$ are retained. In this work, the bath threshold $\eta = 1\cdot 10^{-5}$ was applied uniformly across all fragments and all nine clusters. The resulting embedded clusters span orbital counts (NORB) between 8 and 33 molecular orbitals per fragment, reflecting the local chemistry of each site.

For each fragment, Vayesta writes a self-contained HDF5 archive containing the embedded one-electron Hamiltonian, the embedded two-electron integrals, and the fragment-space and cluster-space molecular-orbital coefficients. These archives are the input to the post-Hartree--Fock solvers and to the partitioned-cumulant energy assembler described below. The EWF fragmentation and MP2 bath construction were performed on the MSU HPCC.

\section{Post-Hartree--Fock Cluster Computations}

Once the embedded cluster Hamiltonians have been written to disk, each fragment is solved at a correlated level to obtain its energy and reduced density matrices for the global reconstruction (Section~\ref{sec:recon}). The solver is chosen according to the size of the embedded cluster, as described next.

\subsection{Solver selection}

\begin{figure}[H]
  \centering
  \includegraphics[width=\linewidth]{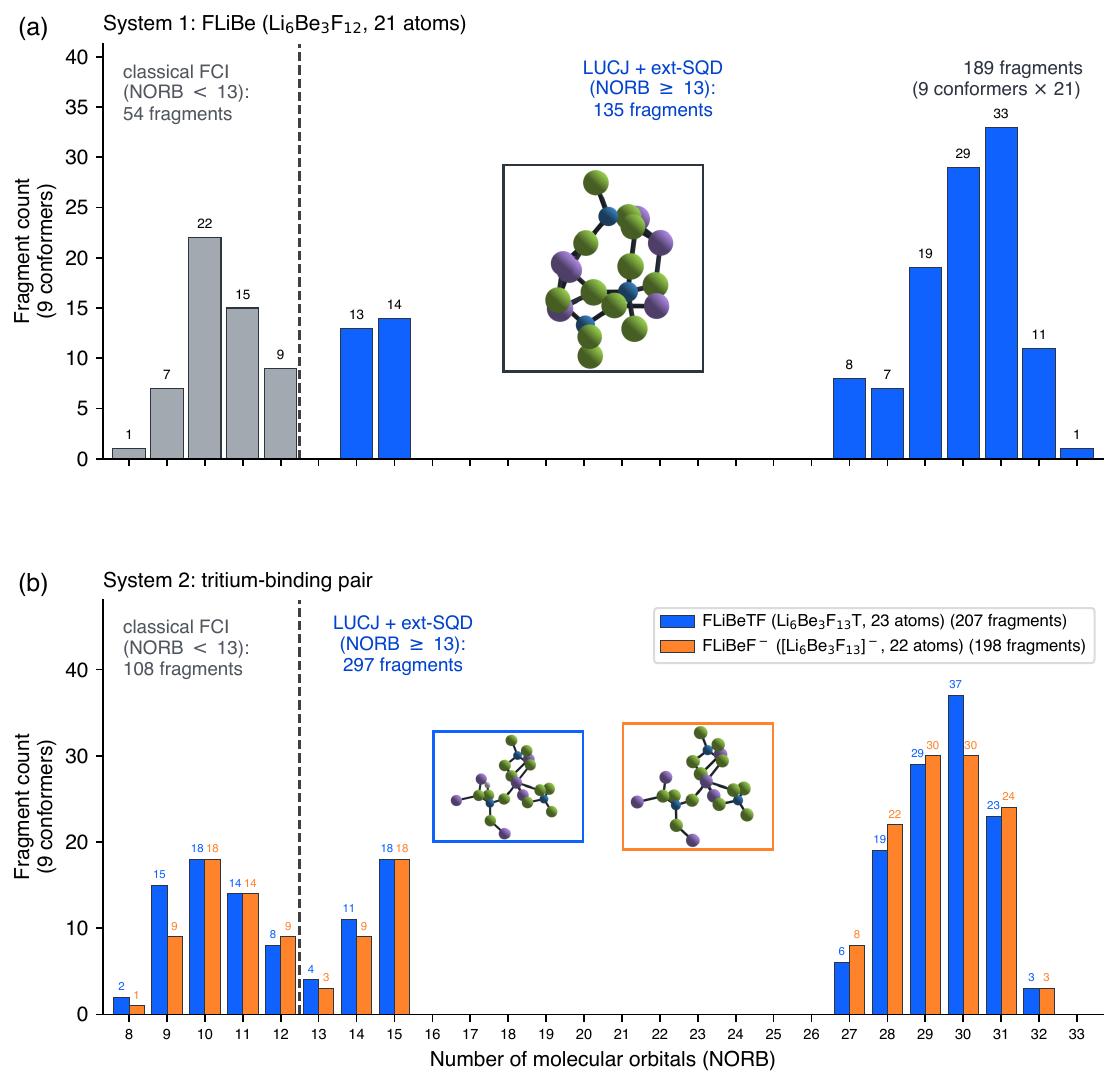}
  \caption{Distribution of embedded-cluster sizes (number of molecular orbitals, NORB) for both benchmark systems.
  \textbf{(a)}~System~1: the 189 EWF fragments of the nine neutral \ce{Li6Be3F12} FLiBe clusters (21 per cluster). The dashed line marks the solver-selection boundary; fragments with $\mathrm{NORB} < 13$ (54 fragments, gray) are solved by classical full configuration interaction, while those with $\mathrm{NORB} \ge 13$ (135 fragments, blue) are dispatched to the IBM Heron-r3 quantum processor and solved with the LUCJ ansatz and extended sample-based quantum diagonalization (ext-SQD).
  \textbf{(b)}~System~2, the tritium-binding pair, across nine clusters each, shown as side-by-side bars per NORB bin: the 23-atom \ce{Li6Be3F13T} cluster (FLiBeTF, blue; 207 fragments) and the 22-atom \ce{[Li6Be3F13]^-} cluster (FLiBeF$^-$, orange; 198 fragments). System-2 fragments follow the same solver routing as System~1 (dashed boundary): those with $\mathrm{NORB} < 13$ (108 fragments) are solved by classical FCI and those with $\mathrm{NORB} \ge 13$ (297 fragments) by LUCJ + ext-SQD.}
  \label{fig:fragsize}
\end{figure}

The 21 fragments of each cluster span a wide range of active-space sizes, and we group them by orbital count to assign a correlated solver. The compact, strongly ionic Li-centred fragments yield small clusters of $\mathrm{NORB} = 8$--$12$, which are solved exactly by classical full configuration interaction (FCI). The more strongly coupled Be--F and bridging-fluoride environments produce larger clusters of $\mathrm{NORB} = 14$--$33$, for which exact diagonalization is feasible for one-off validation but prohibitive at campaign throughput. These are routed to the IBM Heron-r3 quantum processor and treated with extended sample-based quantum diagonalization (ext-SQD). The dispatch boundary is set at $\mathrm{NORB} = 13$: fragments with $\mathrm{NORB} < 13$ are solved classically, while those with $\mathrm{NORB} \ge 13$ are sent to the QPU. The distribution of cluster sizes across the nine clusters is shown in Figure~\ref{fig:fragsize}(a): of the 189 embedded fragments (21 per cluster), 54 fall in the classical-FCI window and 135 in the ext-SQD window.

\subsection{Classical reference solvers: FCI and TCI-8}

The classical embedded references reported in this work---EWF-FCI for System~1 and EWF-FCI+TCI for System~2---require a correlated solution of every fragment that is exact, or numerically indistinguishable from exact, so that any deviation of the quantum-augmented EWF-FCI+ext-SQD total can be attributed to the hardware solver and not to the reference. Full configuration interaction (FCI) was used for all 189 fragments of the nine 21-atom \ce{Li6Be3F12} (System~1) clusters, for all 198 fragments of the 22-atom \ce{[Li6Be3F13]-} anion, and for all but five of the 207 fragments of the 23-atom \ce{Li6Be3F13T} (FLiBeTF) clusters.

The remaining five FLiBeTF fragments have 30 or 31 spatial orbitals with 12 electrons, corresponding to FCI determinant spaces of dimension $3.5\times 10^{11}$ and $5.4\times 10^{11}$, respectively. These cases demand far more resources than any other fragment, typically 480 or more Frontier nodes for 5--6 hours per job. The full FCI campaign for these fragments is being completed; in the interim we use a highly accurate truncated configuration-interaction solver, TCI-8, as their classical reference, and denote the resulting cluster totals EWF-FCI+TCI.

The TCI-8 subspace is built by retaining all $\alpha$ bit-strings within a Hamming distance of at most 8 from the Hartree--Fock reference determinant and forming their tensor product with the identical list of $\beta$ bit-strings. The label TCI-8 refers to this Hamming-distance cutoff of 8 bit positions (equivalently, up to four-fold quadruple excitations within each spin string), not to an energy or wavefunction-coefficient threshold. The resulting subspaces span approximately 12\% of the full Hilbert space for the (30 orbital, 12 electron) fragments and 11\% for the (31 orbital, 12 electron) fragments. To validate the truncation we also completed FCI for five (30 orbital, 12 electron) fragments: the TCI-8 ground-state energies match FCI to well within one microhartree while consuming an order of magnitude fewer computational resources. TCI-8 therefore furnishes a reference of effectively FCI quality for the five large FLiBeTF fragments at a small fraction of the cost.

\subsection{Quantum Circuit Generation: the LUCJ Ansatz}

For each fragment selected for quantum processing --- those with 13 or more spatial orbitals ($\mathrm{NORB} \ge 13$), up to a maximum of 33 --- we generate a parameterized quantum circuit that prepares an approximate ground state of the embedded cluster Hamiltonian. We start from the embedded second-quantized Hamiltonian of the fragment and map fermions to qubits via the Jordan--Wigner transformation, yielding two qubits per spatial orbital (alpha and beta) so that the largest fragments map to 66-qubit problems.

We prepare our quantum circuits using a truncated version of the local unitary cluster Jastrow (LUCJ) ansatz~\cite{motta2023bridging,sung2026ffsim},
\begin{equation}
|\Psi\rangle \;=\; \prod_{\mu=0}^{L-1} e^{\hat{K}_{\mu}}\, e^{i\hat{J}_{\mu}}\, e^{-\hat{K}_{\mu}}\, |x_{\mathrm{RHF}}\rangle,
\end{equation}
where 
\begin{equation}
\hat{K}_{\mu} = \sum_{pr,\sigma} K^{\mu}_{pr}\, \hat{a}^{\dagger}_{p\sigma}\hat{a}_{r\sigma}
\;,\;
\hat{J}_{\mu} = \sum_{pr,\sigma\tau} J^{\mu}_{p\sigma,r\tau}\, \hat{n}_{p\sigma}\hat{n}_{r\tau}
\end{equation}
are anti-Hermitian one-body (orbital-rotation) operators and diagonal density--density (Jastrow) operators respectively, and $|x_{\mathrm{RHF}}\rangle$ is the closed-shell restricted Hartree--Fock state expressed in the embedded-cluster orbitals. We use a single layer, $L = n_{\mathrm{reps}} = 1$, which keeps the two-qubit gate budget compatible with the coherence window of the current Heron processor while still admitting an explicit CCSD seed.

The Jastrow operator $\hat{J}_{\mu}$ is restricted to a hardware-respecting locality pattern. Same-spin density--density couplings are placed on nearest-neighbour orbital pairs $(p, p+1)$ for $p = 0, \ldots, n_{\mathrm{orb}}-2$, matching the heavy-hex nearest-neighbour topology of \device{boston}. Opposite-spin density--density couplings are placed on diagonal pairs $(p, p)$ with $p \in \{0, 4, 8, \ldots\}$ up to $n_{\mathrm{orb}}$, i.e.\ on every fourth spatial orbital. An early version of the opposite-spin stride spanned only the first 13 orbitals, inherited from a 13-orbital prototype, which silently under-coupled the two spin sectors for fragments with $n_{\mathrm{orb}} > 13$. The stride was extended to span all $n_{\mathrm{orb}}$ orbitals early in the production campaign, and all reported quantum runs use the corrected coupling pattern.

Parameter initialization is seeded from a classical CCSD calculation on the embedded cluster: the converged CCSD $T_1$ amplitudes initialize the one-body rotation parameters of $\hat{K}_{\mu}$, and the diagonal projections of the $T_2$ amplitudes initialize the density--density couplings of $\hat{J}_{\mu}$. The seeded parameters are then refined by minimizing the LUCJ-state energy against the embedded cluster Hamiltonian using L-BFGS-B with a maximum of 1000 iterations. Because the embedded Hamiltonian is small enough to diagonalize implicitly during the seed step, this stage is purely classical and produces parameters that already lie close to the variational optimum at $L=1$.

The ffsim library~\cite{sung2026ffsim} provides the LUCJ unitary construction. The resulting circuit is then transpiled by Qiskit~1.1.1~\cite{javadi2024quantum} onto the heavy-hex coupling map of \device{boston} (130-qubit IBM Heron r3) and serialized as a logical-frame QPY payload. For each fragment, the workflow stores the trained LUCJ circuit together with its optimized parameters, the embedded active-space Hamiltonian in FCIDUMP format (for cross-validation against classical FCI and ext-SQD), and the L-BFGS-B optimizer trace. On the MSU HPCC, parameter fitting and dump completes in 10--30~min of wall time per fragment, dominated by the embedded-integral evaluation and the CCSD seed. A representative LUCJ circuit, transpiled and laid out on the heavy-hex lattice of \device{boston} for the largest QPU-dispatched fragment (cluster~4, fragment~15, $\mathrm{NORB}=33$), is shown in the main text.

\subsection{Quantum Sampling and ext-SQD}

Quantum sampling was executed on an IBM Heron r3 quantum processor, ibm\_boston. Fragment size was inferred from the FCIDUMP-derived SCF object generated after EWF fragmentation and circuit generation. Each fragment circuit was transpiled, laid out on hardware, submitted through the IBM Sampler with mitigation settings, and finally its counts saved for later analysis. Sampling was done independently for each fragment geometry in every chemical system (for example FLiBe, FLiBeT/TF). Shots were assigned by fragment size with the orbital cutoff set to 20. If a fragment had less than 20 orbitals, then 100,000 shots were used. Otherwise 1,000,000 shots were used. Fragments utilizing 100,000 shots had a runtime of $28.5 \pm 0.5$ seconds, while fragments utilizing 1,000,000 shots had a runtime of $4\text{m}\,27.0\text{s} \pm 1.0$ seconds.

The qubit layout follows the protocol of Ref.~\cite{shajan2025molecular}: the $\alpha$- and $\beta$-spin orbital registers occupy two parallel zig-zag chains on the heavy-hex lattice of the processor, connected through ancillary qubits that mediate the $\alpha$--$\beta$ couplings of the LUCJ ansatz. Among the available zig-zag connectivity patterns, the layout of each fragment circuit was selected by a heuristic that sums the two-qubit gate errors of all couplers and the readout errors of the qubits in the selected layout. Dynamical decoupling was employed, with an XY4 sequence type selected. Measurement twirling was also utilized. No other error mitigation or suppression techniques were used.

\begin{figure*}[htbp]
\centering
\includegraphics[width=1.0\textwidth]{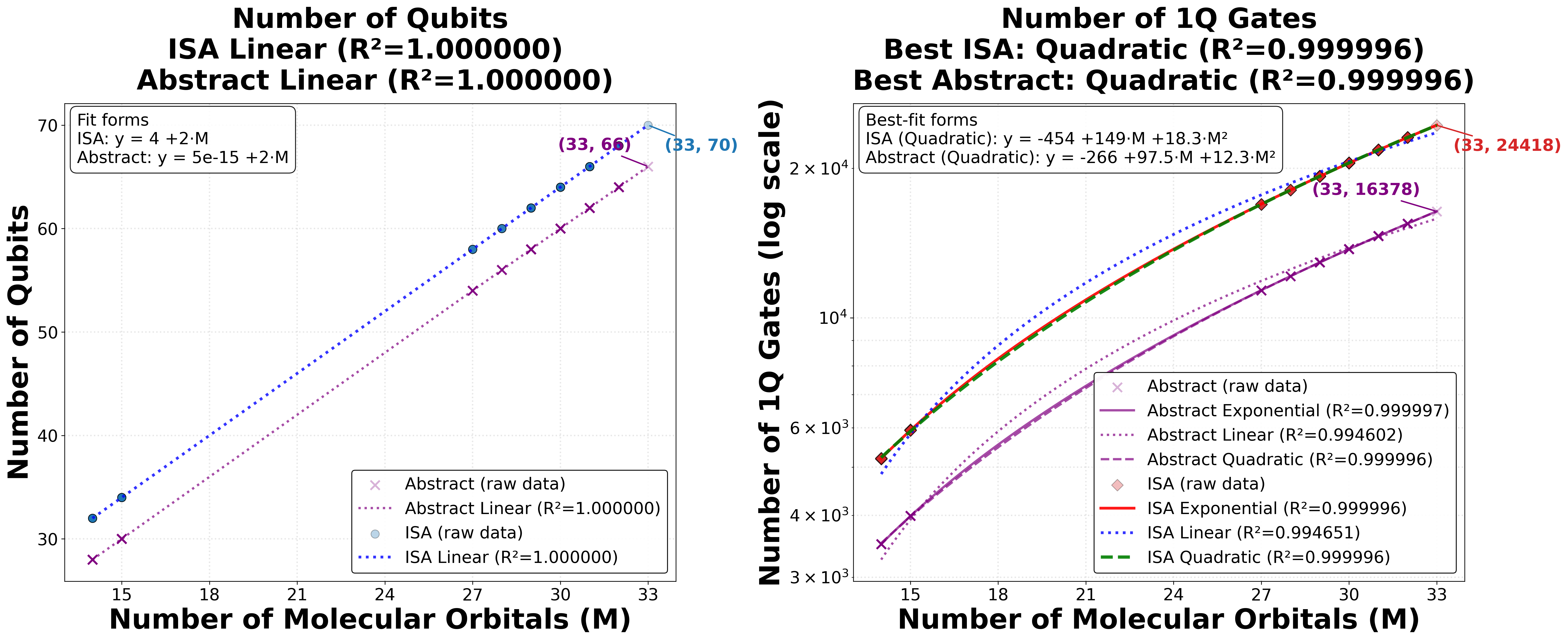}
\caption{Quantum circuit element empirical sub-study. Qubit count scales linearly with molecular orbital count ((M)). The abstract and ISA functional forms are ($y=2M$) and ($y=4+2M$), respectively, with ($R^2=1.000000$) for both fits. One-qubit gate count scales quadratically with molecular orbital count ((M)) at both the abstract and ISA levels. The corresponding functional forms are ($y=-266+97.5M+12.3M^2$) for the abstract circuit and ($y=-454+149M+18.3M^2$) after transpilation, with ($R^2=0.999996$) for both fits.}
\label{fig:circuit_parameter_basic}
\end{figure*}

\textbf{Classical post-processing of quantum samples.} The raw bitstring counts returned by the IBM Sampler primitive are post-processed entirely on classical hardware to produce the extended sample-based quantum diagonalization (ext-SQD) ground-state energy of each QPU-dispatched fragment. The workflow follows the SQD protocol of Refs.~\cite{robledo2025chemistry,motta2024subspace,lassqd} with the configuration-recovery procedure of Ref.~\cite{robledo2025chemistry} and the carryover refinements of Refs.~\cite{shirakawa2025closed,lin2025improved}.

Executing the LUCJ circuit of a fragment on hardware produces a set of measurement outcomes
\begin{equation}
\tilde{\chi} \;=\; \{\mathbf{x} \,\vert\, \mathbf{x} \sim \tilde{p}(\mathbf{x})\},
\end{equation}
in the form of bitstrings $\mathbf{x} \in \{0,1\}^{2M}$, each representing an electronic configuration (Slater determinant) in the embedded-cluster spin-orbital basis. On a noiseless device the configurations would be distributed according to $\vert\langle\mathbf{x}\vert\Psi\rangle\vert^{2}$. On hardware the sampled distribution $\tilde{p}(\mathbf{x})$ deviates from this ideal and, in particular, breaks the conservation of particle number. Samples that violate the symmetries of the embedded Hamiltonian --- total particle number $N$ and the $z$-projection of spin $S_z$ inherited from the closed-shell RHF reference --- are corrected by self-consistent configuration recovery: violating bits are flipped to the closest symmetry-compatible configuration with probabilities weighted by the average orbital occupations, following the procedure of Ref.~\cite{robledo2025chemistry}.

Within each step of configuration recovery, the recovered pool is divided into $K = 10$ batches $\tilde{\chi}_b$ of 3000 samples each; these values are used for all FLiBe fragments. Each batch defines a subspace $S^{(b)}$ spanned by its configurations, into which the embedded many-electron Hamiltonian is projected,
\begin{equation}
\hat{H}_{S^{(b)}} \;=\; \hat{P}_{S^{(b)}}\, \hat{H}\, \hat{P}_{S^{(b)}},
\qquad
\hat{P}_{S^{(b)}} \;=\; \sum_{\mathbf{x} \in S^{(b)}} \vert\mathbf{x}\rangle\langle\mathbf{x}\vert .
\end{equation}
Each projected Hamiltonian is diagonalized exactly using the selected basis diagonalization (SBD) library~\cite{rccs2025sbd}, a parallel sparse exact-diagonalization code, following the integration of SBD within the EWF workflow established in Ref.~\cite{shajan2025molecular}. This yields the batch ground states $\vert\psi^{(b)}\rangle$ and energies $E^{(b)}$, and the lowest energy across the batches, $\min_b E^{(b)}$, is taken as the current estimate of the fragment ground-state energy. The batch ground states are then used to update the average orbital occupations,
\begin{equation}
n_{p\sigma} \;=\; \frac{1}{K} \sum_{b=1}^{K} \langle \psi^{(b)} \vert\, \hat{n}_{p\sigma} \,\vert \psi^{(b)} \rangle ,
\end{equation}
which seed the configuration recovery of the next iteration. The most relevant configurations of each ground-state vector, identified by a squared CI coefficient above $10^{-4}$, are carried over and included in all batches of the next iteration~\cite{lin2025improved}. This carryover ensures that the cumulative subspace converges toward the variationally optimal selected-CI subspace rather than a random subset of the sampled determinants. The self-consistent recovery loop terminates when the energy and the average orbital occupancies change by less than $10^{-8}$~Ha and $10^{-5}$, respectively, or after a maximum of five recovery iterations. With the fixed batch count, this amounts to 50 subspace diagonalizations per fragment. All SQD and ext-SQD calculations were performed on the MSU HPCC using the CPU implementation of the SBD selected-basis diagonalization solver.

\begin{figure}[H]
  \centering
  \includegraphics[width=0.92\linewidth]{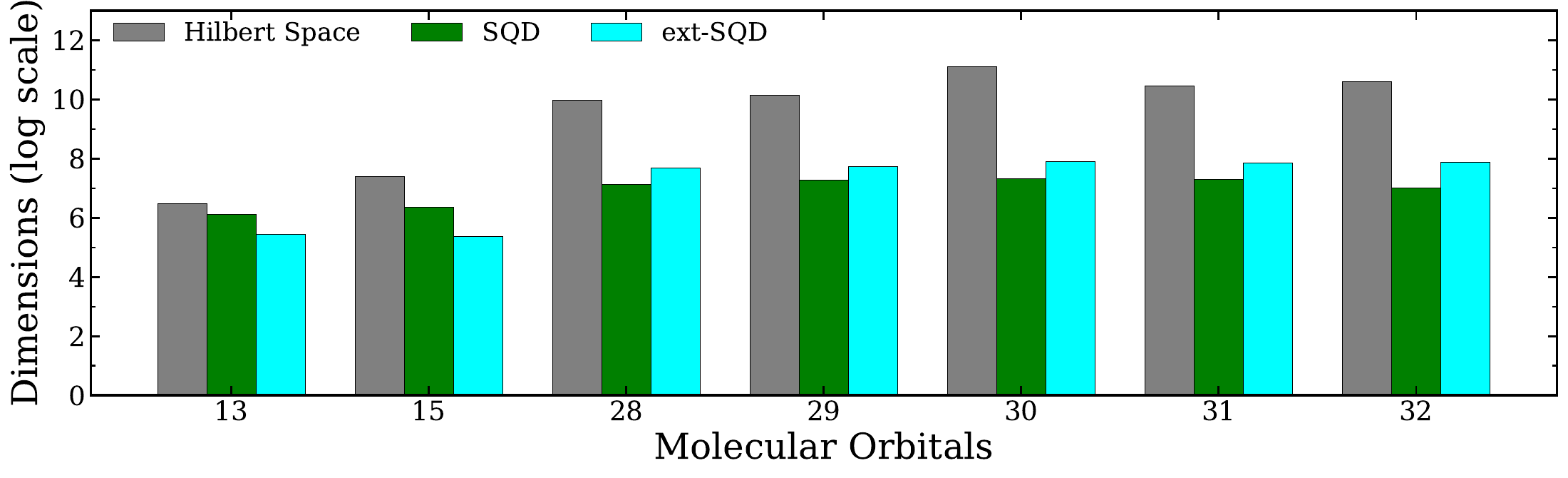}
  \caption{Subspace dimension (logarithmic scale) of the quantum-processor-dispatched embedded fragments as a function of cluster orbital count (NORB), for a representative cluster conformation (cluster~3 of the 23-atom \ce{Li6Be3F13T} set). For each NORB the full configuration-interaction Hilbert-space dimension, $C(\mathrm{NORB},N_\alpha)\times C(\mathrm{NORB},N_\beta)$ (gray), is compared with the SQD subspace (green, averaged over the sampled batches) and the ext-SQD subspace obtained after single particle--hole augmentation (cyan). Where several fragments share the same NORB, their dimensions are averaged. The SQD and ext-SQD subspaces that are explicitly diagonalized are smaller than the full Hilbert space by up to roughly four orders of magnitude for the largest fragments.}
  \label{fig:hilbert}
\end{figure}

Once the recovery loop has terminated, we apply the ``extended'' step of ext-SQD~\cite{barison2025quantum}. The electron configurations are taken from the lowest-energy batch of the final recovery iteration, and the dominant configurations among them are selected by a squared CI coefficient above $10^{-5}$. All single particle--hole excitations from each dominant configuration are then appended to the subspace using the PyCI software package~\cite{richer2024pyci}, growing it by roughly an order of magnitude. The projected Hamiltonian is re-diagonalized in this extended subspace; the resulting energy is the ext-SQD fragment energy reported in the main text. 
The reduction in problem size achieved by this sampling-and-extension procedure is illustrated in Figure~\ref{fig:hilbert} for a representative cluster, which compares the full configuration-interaction Hilbert-space dimension of each QPU-dispatched fragment with the dimensions of the SQD and ext-SQD subspaces that are actually diagonalized.

From the extended ground-state wavefunction we then construct the cluster-space one-body reduced density matrix and the connected two-body cumulant, transform them back into the atomic-orbital basis using the cluster-space molecular-orbital coefficients written by Vayesta, and return them to the EWF partitioned-cumulant assembler. The fragment contributions are then assembled into the EWF-FCI+ext-SQD total of the cluster through the partitioned-cumulant expression of Section~\ref{sec:recon}.

The classical post-processing of the sampled configurations is the dominant classical cost of the quantum workflow. The memory footprint is set by the sparse representation of the projected Hamiltonian and grows steeply with the fragment orbital count, so that the largest NORB~$=33$ fragments are the most demanding to diagonalize. 

The aggregate compute budget of the full campaign, broken down by processor class and system, is summarized in Table~\ref{tab:flibe_resources}. Across all three systems the embedded ext-SQD treatment reproduces the brute-force classical FCI reference at a small fraction of its cost. The contrast is sharpest for the 23-atom \ce{Li6Be3F13T} clusters, where the FCI reference is most demanding (over $2.3\times10^{4}$ GPU node-hours): there ext-SQD requires only a few hundred GPU node-hours, together with fewer than nine QPU-hours, almost two orders of magnitude fewer GPU node-hours than FCI. The full per-method, per-processor breakdown is given in Table~\ref{tab:flibe_resources} and its footnote: Frontier (OLCF) for the FCI reference, the IBM~Heron~r3 processor for quantum sampling, and the MSU and CCF HPC resources for the SQD/ext-SQD classical post-processing.

\begin{table}[htbp]
\centering
\caption{
Net computational resource requirements for HCI, FCI, and SQD+ext-SQD calculations
across all FLiBe-derived systems. QPU time is reported in hours, GPU and CPU usage in
node-hours, and memory as the maximum memory footprint per job.
}
\label{tab:flibe_resources}
\vspace{0.5em}
\setlength{\tabcolsep}{4pt}
\footnotesize
\resizebox{\textwidth}{!}{%
\begin{tabular}{llrrrr}
\toprule
\textbf{System} &
\textbf{Method} &
\textbf{QPU hrs} &
\textbf{GPU node-hrs} &
\textbf{CPU node-hrs} &
\textbf{Max Memory (GiB)} \\
\midrule

\multirow{3}{*}{21 atom FLiBe}& HCI         & 0.00 &     0 & 7.56 &    126 \\
& FCI         & 0.00 &  3416.4 & 0.00 &  16082 \\
& SQD+ext-SQD & 8.91 &  1.14&6870.51*  &     28 \\
\midrule

\multirow{3}{*}{22 atom FLiBeF$^{-}$}& HCI         & 0.00 &     0 & 7.83 &    107 \\
& FCI         & 0.00 &  3055.5 & 0.00 &  11615 \\
& SQD+ext-SQD & 8.66 &  335.8*& 0.00  &     30 \\
\midrule

\multirow{3}{*}{23 atom FLiBeFT}& HCI         & 0.00 &      0 & 8.55 &    136 \\
& FCI         & 0.00 &  23347.8 & 0.00 & 169262 \\
& SQD+ext-SQD & 8.73 &   326.9*& 0.00  &     35 \\
\bottomrule

\end{tabular}}

\vspace{0.5em}

\begin{minipage}{0.95\linewidth}
\footnotesize
HCI calculations were performed on a single node NVIDIA GB200 system with 144 ARM
cores. FCI calculations were performed on Frontier using AMD
MI250X GPUs. The largest FCI calculations utilized up to 480 Frontier nodes
(3840 GPUs). SQD+ext-SQD calculations were performed using IBM Heron r3
processors on \texttt{ibm\_boston}, with supporting classical computation on
NVIDIA A100 or GB200 GPUs and Intel Xeon Platinum 8260 CPUs. \\
* The reported timing was computed by taking one cluster as a representative average of the 9 structures and multiplying across that set of 9.
\end{minipage}

\end{table}

\section{Global Energy Reconstruction}\label{sec:recon}

The fragment solvers return, for every embedded cluster, a correlated ground-state energy together with its one- and two-particle reduced density matrices (RDMs) in the cluster-orbital basis. The total energy of a cluster is reconstructed from these fragment RDMs through the partitioned-cumulant expression of the EWF formalism~\cite{nusspickel2022systematic,nusspickel2023effective}, applied identically regardless of whether a given fragment was solved by classical CCSD, classical FCI, or quantum ext-SQD. The three totals---EWF-CCSD, EWF-FCI, and EWF-FCI+ext-SQD---therefore share the same fragmentation, bath construction, and energy assembly, and differ only in the per-fragment RDMs fed into the reconstruction.

For each fragment the cluster-space RDMs are first transformed into the atomic-orbital basis using the cluster-space molecular-orbital coefficients written by Vayesta, and the resulting contributions are stored as a per-fragment archive. The total electronic energy is then assembled as
\begin{equation}
E_{\mathrm{EWF}} \;=\; E_{\mathrm{RHF}} \;+\; \mathrm{Tr}\!\left[\mathbf{F}\sum_i \Delta\boldsymbol{\gamma}^{(i)}\right] \;+\; \sum_i e^{(i)}_{\mathrm{2c}},
\end{equation}
where $E_{\mathrm{RHF}}$ is the global Hartree--Fock reference energy, $\mathbf{F}$ is the RHF Fock matrix, $\Delta\boldsymbol{\gamma}^{(i)}$ is the correlation correction to the one-body density matrix contributed by fragment $i$ (the partitioned one-body term), and $e^{(i)}_{\mathrm{2c}}$ is the democratically partitioned two-body cumulant contribution of fragment $i$. Each fragment contributes only the share of the cumulant assigned to it by the democratic partitioning of the shared orbital space~\cite{nusspickel2022systematic,nusspickel2023effective}, so that the sum over fragments recovers the global correlation energy without double counting.

In practice the reconstruction proceeds in two stages: a per-fragment stage that converts each solver RDM into its archived contribution, and an aggregation stage that gathers all fragment archives of a cluster, evaluates the expression above, and writes the resulting total energy to a summary table. For the quantum-augmented EWF-FCI+ext-SQD totals reported here, the classical FCI RDMs of the small fragments ($\mathrm{NORB} < 13$) and the ext-SQD RDMs of the large fragments ($\mathrm{NORB} \ge 13$) enter the identical aggregation. The only difference from the fully classical EWF-FCI total is therefore the origin of the large-fragment density matrices. The same partitioned-cumulant reconstruction underlies the protein-scale EWF+ext-SQD study of Ref.~\cite{shajan2025molecular}.

\section{Conformational and Binding Energies}
\label{sec:si_tables}

\begin{figure}[H]
  \centering
  \includegraphics[width=\linewidth]{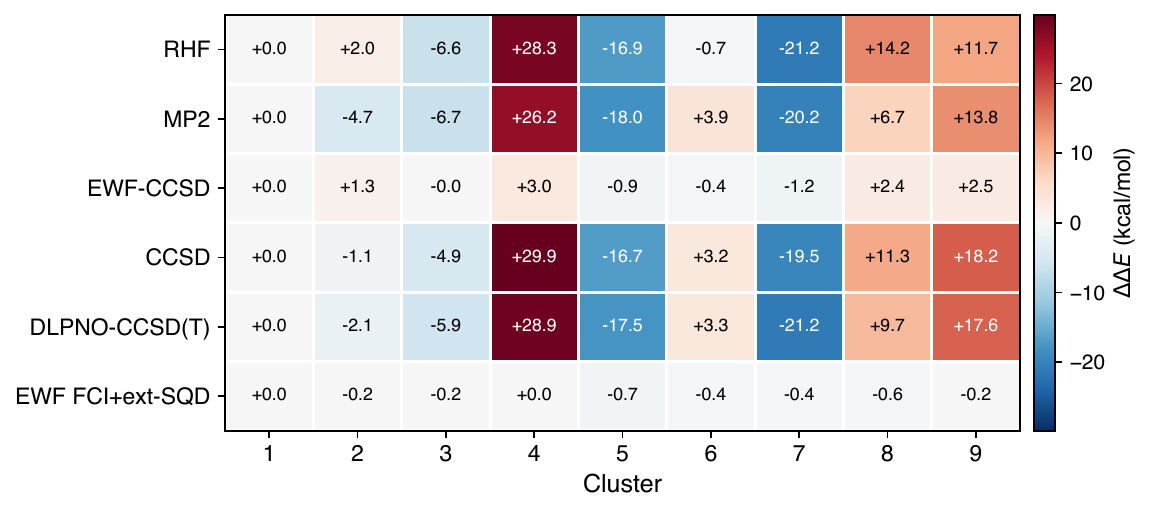}
  \caption{Per-conformation deviation of each method from the embedded-FCI reference, $\Delta\Delta E = \Delta E(\mathrm{method}) - \Delta E(\mathrm{EWF\,FCI})$ (kcal/mol, relative energies referenced to cluster~1), for the nine FLiBe clusters of Table~\ref{tab:absE}. Cell annotations give the signed deviation; the color scale is centered at zero. The conventional single-reference methods (RHF, MP2, canonical CCSD, DLPNO-CCSD(T)) deviate by up to $\approx 30$~kcal/mol (cluster 4), EWF-CCSD remains within 3~kcal/mol, and  EWF-FCI+ext-SQD reproduces its classical embedded-FCI parent to within 0.7~kcal/mol at every cluster conformation.}
  \label{fig:ddEheat}
\end{figure}

Table~\ref{tab:absE} collects the absolute single-point energies of the nine FLiBe clusters at all seven levels of theory, together with the $T_1$ diagnostic of the canonical CCSD calculation. Figure~\ref{fig:ddEheat} resolves the corresponding relative-energy deviations from the embedded-FCI reference, $\Delta\Delta E = \Delta E(\mathrm{method}) - \Delta E(\mathrm{EWF\,FCI})$, cluster by cluster. The  EWF-FCI+ext-SQD row remains below 0.7~kcal/mol in magnitude throughout. The systematic $+2.1$--$2.9$~kcal/mol offset of the hardware energies visible in Table~\ref{tab:absE} cancels in the relative energies.

\begin{table}[H]
  \centering
  \caption{Absolute single-point energies (Hartree) of the nine 21-atom \ce{Li6Be3F12} FLiBe clusters at the 6-31+G* basis, computed at seven levels of theory: restricted Hartree--Fock (RHF), MP2, embedded-wavefunction CCSD (EWF-CCSD), canonical CCSD, DLPNO-CCSD(T), embedded FCI ( EWF-FCI), and the heterogeneous quantum--classical  EWF-FCI+ext-SQD. The final column lists the $T_1$ diagnostic of the canonical CCSD calculation.}

  \label{tab:absE}
  \vspace{0.5em}
  \footnotesize
  \setlength{\tabcolsep}{3.5pt}
  \resizebox{\textwidth}{!}{%
  \begin{tabular}{crrrrrrrc}
    \toprule
    Conf. & \multicolumn{1}{c}{RHF} & \multicolumn{1}{c}{MP2} & \multicolumn{1}{c}{EWF-CCSD} & \multicolumn{1}{c}{CCSD} & \multicolumn{1}{c}{DLPNO-CCSD(T)} & \multicolumn{1}{c}{ EWF-FCI} & \multicolumn{1}{c}{ EWF-FCI+ext-SQD} & $T_1$ \\
    \midrule
    1 & $-1283.138891$ & $-1285.595104$ & $-1284.795064$ & $-1285.599326$ & $-1285.616596$ & $-1284.859026$ & $-1284.854472$ & 0.014 \\
    2 & $-1282.951249$ & $-1285.418144$ & $-1284.608575$ & $-1285.416720$ & $-1285.435493$ & $-1284.674637$ & $-1284.670428$ & 0.015 \\
    3 & $-1283.047862$ & $-1285.504286$ & $-1284.693594$ & $-1285.505617$ & $-1285.524577$ & $-1284.757551$ & $-1284.753250$ & 0.014 \\
    4 & $-1282.741604$ & $-1285.201132$ & $-1284.438086$ & $-1285.199599$ & $-1285.218348$ & $-1284.506873$ & $-1284.502290$ & 0.016 \\
    5 & $-1283.032294$ & $-1285.490272$ & $-1284.662877$ & $-1285.492340$ & $-1285.510926$ & $-1284.725458$ & $-1284.722040$ & 0.014 \\
    6 & $-1283.136032$ & $-1285.584866$ & $-1284.791692$ & $-1285.590283$ & $-1285.607340$ & $-1284.855063$ & $-1284.851085$ & 0.014 \\
    7 & $-1282.945104$ & $-1285.399830$ & $-1284.569431$ & $-1285.402914$ & $-1285.422808$ & $-1284.631514$ & $-1284.627613$ & 0.015 \\
    8 & $-1282.787777$ & $-1285.256053$ & $-1284.462769$ & $-1285.252916$ & $-1285.272732$ & $-1284.530595$ & $-1284.527052$ & 0.016 \\
    9 & $-1282.881653$ & $-1285.334500$ & $-1284.552476$ & $-1285.331690$ & $-1285.349894$ & $-1284.620378$ & $-1284.616124$ & 0.015 \\
    \bottomrule
  \end{tabular}}
\end{table}

Table~\ref{tab:bindE} lists the tritium binding energies of the FLiBe cluster at all eight levels of theory---the values plotted in the main-text binding-energy figure. The five full-molecule methods agree on a strongly bound tritium ion and cluster together, whereas the three embedded EWF methods (which share the same fragmentation and bath) fall a further $\approx 110$~kcal/mol toward weaker binding; within the embedded family,  EWF-FCI+ext-SQD reproduces EWF-FCI+TCI to better than 1~kcal/mol at every cluster conformation.

\begin{table}[H]
  \centering
  \caption{Tritium binding energy $E_{\mathrm{bind}} = E_{\mathrm{FLiBeTF}} - E_{\mathrm{FLiBeF}^-}$ (kcal/mol) of the nine FLiBe clusters at the 6-31+G* basis, at eight levels of theory grouped into full-molecule methods (RHF, PBE-D3, MP2, canonical CCSD, DLPNO-CCSD(T)) and embedded EWF methods (EWF-CCSD,  EWF-FCI+ext-SQD, EWF-FCI+TCI). A dash denotes a non-converged CCSD.}
  \label{tab:bindE}
  \footnotesize
  \setlength{\tabcolsep}{4pt}
  \resizebox{\textwidth}{!}{%
  \begin{tabular}{c rrrrr rrr}
    \toprule
     & \multicolumn{5}{c}{Full-molecule} & \multicolumn{3}{c}{Embedded (EWF)} \\
    \cmidrule(lr){2-6}\cmidrule(lr){7-9}
    Conf. & RHF & PBE-D3 & MP2 & CCSD & DLPNO-CCSD(T) & EWF-CCSD &  EWF-FCI+ext-SQD & EWF-FCI+TCI \\
    \midrule
    1 & $-291.44$ & $-294.03$ & $-285.17$ & $-287.66$ & $-286.83$ & $-183.59$ & $-175.39$ & $-174.84$ \\
    2 & $-379.93$ & $-348.13$ & $-360.89$ & $-359.49$ & $-221.89$ & $-280.75$ & $-282.44$ & $-281.50$ \\
    3 & $-372.45$ & $-363.56$ & $-357.03$ & $-360.90$ & $-359.37$ & $-258.67$ & $-248.87$ & $-248.27$ \\
    4 & $-249.41$ & $-288.40$ & $-244.76$ & --- & $-246.55$ & $-145.72$ & $-138.13$ & $-137.56$ \\
    5 & $-310.48$ & $-314.02$ & $-302.27$ & $-305.25$ & $-304.34$ & $-198.42$ & $-189.20$ & $-188.55$ \\
    6 & $-269.84$ & $-299.40$ & $-265.63$ & $-267.69$ & $-267.13$ & $-164.90$ & $-157.88$ & $-157.20$ \\
    7 & $-317.18$ & $-321.27$ & $-303.64$ & $-307.83$ & $-306.22$ & $-198.48$ & $-188.84$ & $-188.09$ \\
    8 & $-338.85$ & $-333.80$ & $-331.62$ & $-334.06$ & $-333.14$ & $-219.25$ & $-209.97$ & $-209.32$ \\
    9 & $-241.43$ & $-281.13$ & $-237.68$ & $-239.71$ & $-239.33$ & $-141.36$ & $-134.76$ & $-134.14$ \\
    \bottomrule
  \end{tabular}}
\end{table}

\bibliography{ms}
\end{document}